\documentclass[12pt]{article}
\usepackage{graphicx}
\usepackage{booktabs}
 \usepackage{lscape}
 \usepackage{verbatim}
 \usepackage{multirow}
 \usepackage{geometry}
 \usepackage{amssymb}
 \usepackage{units}
 \usepackage{setspace}
 \usepackage{dsfont}
  \usepackage{natbib}
   \usepackage{bm}
   \usepackage{color}
   \usepackage{amsmath}
   \usepackage{comment}
   \usepackage{graphicx,wrapfig,lipsum,multirow,subfigure}

\title{Dose Finding Studies for Therapies with Late-Onset Toxicities: A Comparison Study of Designs}

\author{Helen Barnett$^1$, Oliver Boix$^2$, Dimintris Kontos$^3$, Thomas Jaki$^{1,4}$\\$^1$ MRC Biostatistics Unit, University of Cambridge\\
$^2$ Bayer AG\\
$^3$ ClinBAY\\
$^4$ Department of Mathematics and Statistics, Lancaster University}
\begin{document}
\maketitle
\begin{abstract}
An objective of phase I dose-finding trials is to find the maximum tolerated dose; the dose with a particular risk of toxicity. Frequently, this risk is assessed across the first cycle of therapy. However, in oncology, a course of treatment frequently consists of multiple cycles of therapy. In many cases, the overall risk of toxicity for a given treatment is not fully encapsulated by observations from the first cycle, and hence it is advantageous to include toxicity outcomes from later cycles in phase I trials. Extending the follow up period in a trial naturally extends the total length of the trial which is undesirable. We present a comparison of eight methods that incorporate late onset toxicities whilst not extensively extending the trial length. We conduct simulation studies over a number of scenarios and in two settings; the first setting with minimal stopping rules and the second setting with a full set of standard stopping rules expected in such a dose finding study. We find that the model-based approaches in general outperform the model-assisted approaches, with an Interval Censored approach and a modified version of the Time-to-Event Continual Reassessment Method giving the most promising overall performance in terms of correct selections and trial length. Further recommendations are made for the implementation of such methods.
\end{abstract}

\section{Introduction}

In phase I dose finding studies, the aim is to find the Maximum Tolerated Dose (MTD) to recommend for phase II, defined as the highest dose giving an acceptable level of toxicity \citep{Storer2001}. An acceptable level of toxicity is in general equated to a certain probability of occurrence of a Dose Limiting Toxicity event (DLT), which in oncology is most often defined as a grade 3 or higher toxicity by the grading scale of the National Cancer Institute \citep{NCI}. This probability is usually set to 20-30\% within the general treatment population, with the actual choice of target probability depending on the indication, available treatment option and expected benefit within the target population.

However, this probability refers to the risk of DLT within the follow-up period of the trial, which is typically only one cycle of treatment. In many cancer treatments, the full course of treatment actually consists of multiple cycles of therapy, given sequentially. It is therefore important to consider multiple cycles of therapy in the dose-finding trial. In a review of 445 patients in 36 phase I trials by Postel-Vinay et al. \cite{Postel-Vinay2011}, it was found that 57\% of grade 3 or 4 toxicities occurred after the first cycle of treatment, and that for 50\% of patients, their worst grade of toxicity was observed after the first cycle. This is a clear indication that later cycles are important to include in phase I dose-finding trials, since their omission may lead to missing large amounts of information on toxicity risk of the investigated doses and hence the recommendation of sub-optimal doses. However, by increasing the follow up period, one also greatly increases the trial duration if the entire follow up period must be completed for the previous cohort of patients before the dose for the next cohort can be assigned. Such an increase in trial duration is obviously undesirable, as the focus in such trials is in efficient decision making.

A small number of statistical methods for the design of trials in such a setting have been proposed, with varying approaches to include the later onset toxicities without substantially increasing the trial duration. For example, the Rolling 6 design \citep{Skolnik2008}, a rule based approach that is an extension of the 3+3 \citep[see][]{Storer1989}, uses a set of rules based on the number of DLTs observed and the number of patients who have completed and are yet to complete their full follow up period. Other designs \citep[such as the model-assisted designs by][]{Yuan2018,Guo2019,Lin2020}, follow similar sets of rules, with the addition that escalation is aided by a simple model for the probability of toxicity at each dose level under the assumption of monotonicity, that a higher dose is associated with a higher probability of toxicity. There are also a limited selection of model based designs that account for later onset toxicities \citep{YingKuenCheung2000,Sinclair2014,Doussau2013, Yin2017}. These designs vary in their approach to accounting for the occurrence of DLTs in the different treatment cycles. The Time-to-Event approach of \citep{YingKuenCheung2000} models the DLT occurrence using a time-to-event variable defined on the entire follow up period, and not necessarily breaking down this period into cycles. The other model-based approaches do break the follow up period into the respective cycles, with the interval censored approach of Sinclair and Whitehead \cite{Sinclair2014} modelling the probability of DLT in each cycle, conditional on the lack of DLT in all previous cycles, whereas the method by Doussau et al. \cite{Doussau2013} fits a proportional-odds mixed model to data from the different cycles. The approach by Yin et al. \cite{Yin2017} fits a linear mixed effects model including a cycle effect to data of total toxicity profile, including grades and type of toxicity.

As expected with novel statistical methodology, each approach is praised by the respective authors for its advantages over another given method in any particular setting. However, such settings usually fit well to the approach suggested, and although some exploration of settings that violate assumptions is often undertaken, it would be of great aid to have a comparison of the leading methods in settings that are both realistic and not adhering to the assumptions of the approaches. Therefore in this work we undertake a simulation study to compare the most prominent methods for dose-finding studies incorporating late-onset toxicities, in order to evaluate the strengths and weaknesses of each of the methods, and their applicability to phase I dose finding studies. In this comparison we use modified versions of the methods, to improve their applicability in this setting and to ensure their comparability.

This paper has the following structure. In Section \ref{sec:methods}, the eight approaches for comparison are outlined, with notations introduced and assumptions of each method highlighted. In Section~\ref{sec:setting}, we introduce the setting for the simulation study, and in Section~\ref{sec:prior} we describe the procedure we use to choose the values for the hyper-parameters of the prior distributions in order for a fair comparison across methods. In Section~\ref{sec:simulations}, we present the results of the simulation study, before concluding with a discussion in Section~\ref{sec:discussion}.

\section{Methods} \label{sec:methods}
In this section we outline the eight methods that are implemented in the comparison study. The purpose is to give an overview of each method, with key details on the different models used in the dose escalation. Further and more in-depth descriptions can be found in the relevant referenced literature. We highlight any modifications made to the original proposals.

In each method, the following notation is consistent throughout. Consider a dose-finding study where $J$ dose levels labelled {$d_j$} for {$j=1, \ldots , J$} are investigated. Each patient {$i$} is followed up for {$S$} cycles of therapy indexed {$s= 1, \ldots , S$}. A new cohort of patients is admitted at the beginning of each cycle, so that partial information is available for the previous {$S-1$} cohorts when assigning the dose for the current cohort. For example, for a follow up period of 3 cycles, partial information is available for the previous 2 cohorts; information is available for only the first cycle of therapy of the previous cohort, and information on the first two cycles for the second previous cohort.
Dose assignment is based on some target, {$\tau$}, defined as the probability of observing a DLT in the entire follow-up of $S$ cycles. 

The general process of a dose-finding study is as follows. Pre-trial, for a model-based method, a dose response model is chosen and prior distributions assigned to the parameters. For a model-assisted method, prior distributions are assigned to  any relevant parameters that guide escalation, and a decision table based on these and all possible outcomes is calculated. The first cohort of patients is assigned to the lowest dose. After one cycle of treatment, the observed responses from these patients are used to decide which dose to assign to the next cohort of patients to, or to stop the trial. For a model-based method the posterior distribution is updated from which the next best dose is derived and for a model-assisted method the result is looked up in the decision table. This same process is repeated after each cycle of treatment, until the trial is stopped for a prespecified reason, and either a dose is recommended as the MTD or all doses are deemed unsafe in which case no MTD is recommended.

\subsection{Time-to-event Continual Reassessment Method Approach (TITE-CRM)}
The first method we review is perhaps the most well known, the Time-to-event Continual Reassessment Method Approach (TITE-CRM) first proposed by Cheung and Chappell \cite{YingKuenCheung2000}.  We consider two approaches under the umbrella of TITE-CRM, an approach closely mirroring the original proposed methodology by Cheung and Chappell \cite{YingKuenCheung2000} (1-Parameter TITECRM) and a modification to include the actual dose values instead of standardized doses (2-Parameter TITECRM).

\subsubsection{1-Parameter TITECRM}
Here, a weighted dose response model is used:
\[ 
{G(d,w,\beta) = w F(d,\beta)},
\]
where {$0 \leq w \leq 1$} is a function of time-to-event of a patient response, $F$ is the assumed dose-response model, $d$ is the scaled dose and $\beta$ is the parameter to be estimated. The scaled doses are interpreted as the prior belief of the probability of toxicity on that dose, these are used as opposed to real values of the dosages. The dose-response model suggested is $F(d,\beta)= d^{\hspace{2pt} \mbox{\scriptsize exp}(\beta)}$ and a Normal prior distribution is elicited on $\beta$ ($\sim N(0,\sigma^2)$), with the posterior distribution updated after each cycle using likelihood
\[
{\mathcal{L}(\beta) = \prod_{i=1}^{n} G( d_{[i]}, w_{i,n}, \beta)^{y^{(TC)}_{i,n}} \{ 1-G( d_{[i]}, w_{i,n}, \beta)\}^{1-y^{(TC)}_{i,n}}},
\]
where $n$ is the number of patients treated, and $y^{(TC)}_{i,n}$ is an indicator taking the value 1 if patient $i$ has observed a DLT after information is available for at least one cycle for $n$ patients.
Dose assignment is determined by minimising $|F(d_j, \hat{\beta_n}) - \tau |$ where $\hat{\beta_n}$ is the posterior mean of $\beta$ after information is available for at least one cycle for $n$ patients and $\tau$ is the target $P(DLT)$ for all $S$ cycles. We use the weights suggested by Cheung and Chappell \cite{YingKuenCheung2000}, of the simple $w_{i,n}=u_{i,n}/S$, where $u_{i,n}$ is the current number of cycles patient $i$ has been observed for. As outlined by Cheung and Chappell \cite{YingKuenCheung2000}, if a DLT is observed then $w_{i,n}=1$. The final dose recommendation is the dose level that minimises $|F(d_j, \hat{\beta_n}) - \tau |$ once the follow-up for all enrolled patients has completed.

Although this method is flexible enough to allow for continuous time-to-event responses, here we discretize this variable according to the cycle the response is observed in. The TITE-CRM has an initial period where the dose is escalated one level at a time until a DLT is observed. In the original methodology, in this initial period, each patient is followed up for their entire follow up time before the next patient's dose is assigned. In our implementation, only one cycle is required for follow up before the next is assigned, in line with the rest of the trial.
\subsubsection{2-Parameter TITECRM}
In order to make use of the real doses used in the trial, we modify the TITE-CRM to include these, henceforth labelled as the TITE-CRM2. Since the above dose-response model is only valid for $0<d<1$ we now use the 2-parameter logistic model:
\[
F(d_j,\mathbf{\beta})= \frac{\mbox{exp}(a_0 + a_1 d_j)}{1+\mbox{exp}(a_0 + a_1 d_j)},
\]
where $d_j$ is the real value of the dose.
Here, MCMC methods must be used to update the posterior distribution for $\bm{\beta}=(a_0, a_1)$. The prior distributions are $a_0 \sim N(\mu_{a_0},\sigma^2_{a_0})$ and  $log(a_1) \sim N(\mu_{a_1},\sigma^2_{a_1})$. These Normal priors are in line with other dose-finding methods \citep[e.g.][]{Neuenschwander2008}.

\subsection{Interval Censored Survival Approach (ICSDP)}

This approach introduced by Sinclair and Whitehead\cite{Sinclair2014} uses $\pi_{j,s}$, the conditional probability of observing a DLT in cycle $s$ for a patient on dose $d_j$ given they did not observe a DLT in previous cycles. The prior for this method takes the form of pseudo-data, based on the approach by Whitehead and Williamson \cite{Whitehead1998}, where a small number of pseudo-patient observations (allowing for non-integer observations) used to guide the escalation. Here we have $n_0$ pseudo-patient observations on dose $d_1$, with $\pi^*_1 n_0$ patients observing a DLT on the first cycle and $n_0$ pseudo-patient observations on dose $d_J$, with $\pi^*_J n_0$ patients observing a DLT on the first cycle. Pseudo-data for subsequent cycles is calculated based on a decreasing $\pi^*_j$.

With the pseudo-data prior, the posterior for $\theta$ and $\bm{\gamma}$ is updated using likelihood:
\[
{\mathcal{L}(\theta, \bm{\gamma}) = \prod_{j=1}^{J} \prod_{s=1}^{S} \pi_{j,s}^{r_{j,s}} (1-\pi_{j,s})^{q_{j,s}}},
\]
with link function $\log (-\log(1-\pi_{j,s})) = \gamma_s + \theta \log(d_j)$, where {$r_{j,s}$} is the number of patients that experience a DLT in cycle $s$, $q_{j,s}$ is the number of patients who have completed $s$ cycles without experiencing a DLT, $\gamma_s$ is an intercept term for cycle $s$.
Dose assignment is determined by maximising gain function $1/(\tau - \hat{\rho_j}(t_S))^2$ where $\tau$ is the target $P(DLT)$ for all $S$ cycles and $\hat{\rho_j}(t_S)$ is the current estimate of $P(DLT)$ for dose $d_j$ for all $S$ cycles. The final dose recommendation is the dose level that maximises $1/(\tau - \hat{\rho_j}(t_S))^2$  once the follow-up for all enrolled patients has completed. This is a modification from the original proposal, which recommended any dose on the continuous scale, which we make in order for results of this design be measured by the same metric as other designs.

\subsection{Proportional Odds Mixed Effect Model Approach (POMM)}

The third approach is the Proportional Odds Mixed Effect Model Approach (POMM) proposed by Doussau et al. \cite{Doussau2013}. This method uses the additional information on the grades of toxicity observed in each cycle. Due to difficulties in model fitting in the original proposal, we make two modifications to improve the stability of the method. We introduce the use of prior information to align this method with the other Bayesian approaches, as well as to aid the model fitting. This prior is in the form of pseudo-data, with responses from pseudo-patients on all doses and cycles, down-weighted so as to not outweigh the observed data. We also alter the target probability of DLT to include all cycles, as opposed to per cycle. Again, this aligns with other methods but also allows for differing risks across cycles.

Toxicities are categorised by grade, the response variable for subject $i$ in cycle $s$ is defined as:
\begin{equation*}
{
  y^{(POMM)}_{i,s} =
    \begin{cases}
      1 & \text{if no toxicity or grade 1 toxicity},\\
      2 & \text{if grade 2 toxicity}, or\\
      3 & \text{if grade 3+ toxicity}.
    \end{cases}       
    }
\end{equation*}

For the first 15 subjects, the following generalised linear model is used:
\[
logit(P(Y^{(POMM)}_{i,1}=3)|d_j) = b_0 + b_1 d_j ,
\]
so that there are no mixed effects and only the responses from cycle 1 are used. Dose assignment is determined by minimising {$|P(Y^{(POMM)}_{i,1}=3|data,\hat{b_0},\hat{b_1},d_j) -\tau_{s=1}|$} where {$\tau_{s=1}$} is the target $P(DLT)$ for cycle 1.

From subject 16 onwards, a logistic proportional odds mixed-effect regression model is used
\[
{
logit(P(Y^{(POMM)}_{i,s} \leq k | d_j)) = \alpha_k - \beta_1 d_j - \beta_2 s - u_i \hspace{5pt} \mbox{for} \hspace{5pt} k=1,2,
}
\] 
where $u_i \sim N(0, {\sigma_0}^2)$ is the individual patient effect. $\bm{\theta}= (\alpha_1,\alpha_2, \beta_1, \beta_2)$.
Dose assignment is determined by minimising {$|P(Y^{(POMM)}_{i,-}=3|data,\hat{\bm{\theta}},d_j,u_i=0) -\tau|$} where {$\tau$} is the target $P(DLT)$ for the whole follow up period and $P(Y^{(POMM)}_{i,-})$ is the estimated probability of DLT in the whole follow up period for a patient with $u_i=0$. 

The final dose recommendation is the dose level that minimises {$|P(Y^{(POMM)}_{i,-}=3|data,\hat{\bm{\theta}},d_j,u_i=0) -\tau|$} (or {$|P(Y^{(POMM)}_{i,1}=3|data,\hat{b_0},\hat{b_1},d_j) -\tau_{s=1}|$} if there are less than 16 patients) once the follow-up for all enrolled patients has completed.

\subsection{Total Toxicity Profile Approach (nTTP)}

This approach proposed by Yin et al. \cite{Yin2017} calculates a normalized Total Toxicity Profile ({nTTP}) value for each patient using information on grades and types of toxicity. The flexibility of this approach allows for any number of types of toxicity, provided the specification for the method of calculating the nTTP value is decided before the trial. Here, following from Yin et al. \cite{Yin2017}, three types of toxicity are included (renal, haematological, neurological), each with grades 0-4. Patients can observe any combination of grades and types of toxicity within a given cycle, and the maximum grade of each type for each cycle is recorded. This is used to calculate the nTTP value using the weights specified by Yin et al. \cite{Yin2017}.
A linear mixed effect model is fitted to the nTTP values for all cycles: 
\[
{
y^{(nTTP)}_{i,s} = \beta_0 + \beta_1 x_i + \beta_2 s + \gamma_i + \epsilon_{i,s}},
\]
where $y^{(nTTP)}_{i,s}$ is the observed nTTP value for patient $i$ on cycle $s$, $x_i$ is the dose assigned to patient $i$, $\gamma_i \sim N(0, {\sigma_{\gamma}}^2)$ is the individual patient effect and $\epsilon_{i,s} \sim N(0, {\sigma_{\epsilon}}^2 )$  is measurement error.
The following priors are elicited: $\beta_0 \sim N(\mu_{\beta_0}, \sigma^2_{\beta_0})$, $\beta_1 \sim N(\mu_{\beta_1}, \sigma^2_{\beta_1})$, $\beta_2 \sim N(\mu_{\beta_2}, \sigma^2_{\beta_2})$, ${\sigma_{\gamma}}^2 \sim IG(0.001,0.001)$  and ${\sigma_{\epsilon}}^2 \sim IG(0.001,0.001)$ and the posterior distributions for all parameters are updated after each cycle.
Dose assignment is determined by minimising $|nTTP(d_j, cycle \hspace{2pt}1) - \tau_{nTTP}|$ where $\tau_{nTTP}$ is the target nTTP for cycle 1 and $nTTP(d_j, cycle\hspace{2pt} 1)$ is the current posterior estimate of nTTP for dose $d_j$ for cycle 1. Although the nTTP value for each cycle is included in the model, only the current posterior estimate of nTTP for cycle 1 is included in the criterion. Because of the nature of the nTTP variable, there is no natural measure across all cycles in the same way there is with a variable of probability and hence we here use the the criterion using only cycle 1 as recommended by Yin et al. \cite{Yin2017}.

The final dose recommendation is the dose level that minimises $|nTTP(d_j, cycle \hspace{2pt}1) - \tau_{nTTP}|$ once the follow-up for all enrolled patients has completed.

\subsection{Time to Event Bayesian Optimal Interval (TITE-BOIN) Approach}
This model-assisted approach proposed by \cite{Yuan2018} is a time-to-event extension of the Bayesian Optimal Interval design, whereby dose escalation is guided by the target interval ($\tau_1$, $\tau_2$). Doses are escalated or de-escalated one dose level at a time, and this decision is based on two metrics: the escalation and de-escalation boundaries of the non-time-to-event versions ($\lambda_d$ \& $\lambda_e$) and the standardized total follow-up time (STFT). A set of rules based on these two values determines whether the dose is escalated, de-escalated or remained constant.  A decision table for these rules can be calculated before the trial, based on the target interval and the Beta($\alpha$, $\beta$) prior assigned to the probability of DLT at each dose. The values of $a$ and $b$ are originally chosen so that the prior has an effective sample size of 1 and mean $\tau/2$, with this prior the same across all doses. The target interval used by the authors is (0.6$\tau$, 1.4$\tau$). The final dose recommendation is based on an isotonic regression once the follow-up for all enrolled patients has completed.
\subsection{Time to Event Modified Toxicity Probability Interval (TITE-mTPI2)} 
This model-assisted approach by \cite{Lin2020} is a time to event extension of the Modified Toxicity Probability Interval 2 design also known as the `Keyboard' design, whereby dose escalation is guided by the target interval ($\tau_1$, $\tau_2$) = ($\tau-\epsilon_1$,  $\tau+\epsilon_2$) where $\tau$ is the target probability of DLT. The interval [0,1] is divided into equally sized ``keys'' of size $\epsilon_1 + \epsilon_2$ (apart from the end keys) and the key that has the largest probability guides whether the dose is escalated or de-escalated. These probabilities are calculated based on ``effective'' binomial data, including the patients whose full observation period has not yet completed. Again, the decision table can be calculated before the beginning of the trial, based on the target interval and the Beta(1, 1) prior assigned to the probability of DLT at each dose. The target interval used by the authors is ($\tau-0.05$, $\tau$+0.05). The final dose recommendation is based on an isotonic regression once the follow-up for all enrolled patients has completed.
\subsection{Rolling Modified Toxicity Probability Interval (R-mTPI2)}
This model-assisted approach introduced by \cite{Guo2019} is an extension the rolling 6 design and of the Modified Toxicity Probability Interval design, whereby dose escalation is guided by the target interval ($\tau_1$, $\tau_2$) = ($\tau-\epsilon_1$,  $\tau+\epsilon_2$) where $\tau$ is the target probability of DLT. Again, the interval [0,1] is divided into equally sized ``keys'' of size $\epsilon_1 + \epsilon_2$ (apart from the end keys). In the traditional mTPI2 design, the key that has the largest probability guides whether the dose is escalated or de-escalated, whereas in this rolling version, the decision to escalate is based on a series of rules. These rules first consider the escalation decision based solely on the number of observed DLTs within completely observed patients, then consider decisions based on the best/worst case scenarios that none/all incomplete observations are DLTs, and finally a consideration of how many patients have been consecutively assigned to the current dose without interruption. Like the other two model-assisted methods, a decision table can be calculated before the start of the trial, based on the target interval and the Beta(1, 1) prior assigned to the probability of DLT at each dose. The target interval used by the authors is ($\tau-0.05$, $\tau$+0.05). The final dose recommendation is based on an isotonic regression once the follow-up for all enrolled patients has completed.

\section{Setting} \label{sec:setting}
We consider the motivating example of the phase I trial, `First-in-human Study of BAY2287411 Injection, a Thorium-227 labelled Antibody-chelator conjugate, in patients With tumours known to express mesothelin' \citep{Example_trial1}, an ongoing trial which started in June 2018. Subjects with either advanced recurrent epithelioid mesothelioma or serous ovarian cancer who have exhausted available therapeutic options are given a single dose of Thorium-227 on Day 1 of each cycle lasting 6 weeks. The dose starts at 1.5 MBq and increases in steps of 1.0 or 1.5 MBq.

In this example trial, the follow-up for observation of DLTs is only the 1st cycle of treatment, up to day 43, with a target DLT rate of 0.3. We base the setting for our simulation study on the motivating example, exploring the impact of considering later onset toxicities in the design. Six doses of therapy are investigated, of quantity 1.5MBq , 2.5MBq , 3.5MBq , 4.5MBq , 6.0MBq , 7.0MBq. The study enrols a maximum of 30 patients in cohorts of size 3. Patients are followed up for a total of 3 treatment cycles, each of length 6 weeks. If a DLT response is observed, that patient goes off study. The first cohort of patients are always enrolled at the lowest dose.

In order to review the performances, we consider two settings, the first with minimal stopping rules, and the second with a set of stopping rules that are used as standard in such studies in practice.  The following enforcement and stopping rules are considered, although individual numbers can be adapted according to the study itself. For any given dose, $p_s$ is the $P(DLT)$ in cycles up to and including cycle $s$.\\

\textbf{\textit{Enforcement Rules:}}
\begin{enumerate}
\item \textbf{Hard Safety}: If there is a high probability that the dose exceeds the target, that dose and all above is excluded from further experimentation. (i.e. A dose is excluded when $P(p_1>\tau)>threshold$) Here we use a threshold for excessive toxicity of 95\%, with a Beta(1,1) prior, which translates to the following numbers. For any given dose, in the first cycle, if there are at least 3 DLT responses out of 3 patients, or at least 4 DLT responses out of 6 patients, or at least 5 DLT responses out of 9 patients, then all dose assignments must be lower than that dose for the rest of the study . (i.e. A dose is excluded when $P(p_1>30\%)>0.95$)
\item \textbf{K-fold Skipping Doses}: No more than a 2-fold-rise in dose based on the highest experimented dose so far.
\end{enumerate}

\textbf{\textit{Stopping Rules:}}
\begin{enumerate}
\item \textbf{Sufficient Information}: If a dose is recommended on which 9 patients have already been treated, the trial is stopped.
\item \textbf{Lowest Dose Deemed Unsafe}: $P(p_1>30\%)>0.80$ for dose $d_1$ and at least one cohort of patients has been assigned to dose $d_1$.
\item \textbf{Highest Dose Deemed Very Safe}: $P(p_1\leq 30\%)>0.80$ for dose $d_J$ and at least one cohort of patients has been assigned to dose $d_J$.
\item \textbf{Precision}: Stopping when MTD is precisely estimated, $CV(MTD)<30\%$. The coefficient of variation is calculated as an adjusted median absolute deviation divided by the median. This stopping rule is only used once at least 9 patients have at least one cycle of treatment (on any dose).
\item \textbf{Hard Safety}: The lowest dose is considered unsafe according to the hard safety enforcement rule.
\item \textbf{Maximum Patients}: The maximum number of patients ($N=30$) have been recruited.
\end{enumerate}

In setting 1, we only consider the enforcement rule of no k-fold dose skipping, and the stopping rules of sufficient information and maximum patients. In setting 2, all enforcement and stopping rules are applied. Note that the model-assisted methods are unable to stop for precision since they only consider discrete dose levels with no model relating dose value and response. Although stopping rules 2 and 5 both stop the trial for safety concerns, it is important to highlight that they do so in a different manner. The hard safety stopping rule only considers the lowest dose in isolation, and does not use the analysis from the design itself. The lowest dose deemed unsafe rule uses the full observed data and the method of analysis from the design itself. Stopping rules 2 and 3 are implemented for the model assisted designs using the assisting Beta-binomial model, and for TITECRM and TITECRM-2 by using an additional model which restricts the total follow up time to the length of one cycle.

\section{Prior Calibration} \label{sec:prior}
The value of hyper-parameters of the prior distributions can have a substantial effect on the dose escalation. In a clinical setting, these can reflect belief of the toxicity of the doses, but they also have a key role in the safe and controlled escalation procedure. In order for a fair comparison between these methods, we use a calibration procedure, in line with that used by Mozgunov et al. \cite{Mozgunov2021}, where further details to supplement the outline we present here can be found.

This calibration procedure is conducted as follows. For any given design, a grid search is performed over values of the hyper-parameters in order to find the combination of hyper-parameter values that gives the best overall performance. This performance is measured as the geometric mean of proportion of correct recommendations of MTD in 1000 simulations across a small set of clinically plausible settings. The geometric mean is used to penalise a very poor performance in any given scenario. This calibration procedure gives each design the same opportunity to achieve a good performance. 

The priors are calibrated separately for setting 1 and setting 2, as setting 2 includes safety stopping rules so we must consider performance in scenarios where all doses are unsafe and where all doses are too safe. In setting 1, we cannot calibrate using such scenarios as there is no `correct' outcome.

The following scenarios in Table~\ref{tab:prior_sc} are considered for prior calibration. In setting 1, P.S.1-P.S.4 are used, and in setting 2 we introduce P.S.5 and P.S.6, to reflect the addition of the stopping rules. In scenario P.S.5, a `correct' outcome is stopping according to stopping rules 2 or 5, not recommending any dose. In scenario P.S.6, a `correct' outcome is stopping according to stopping rule 3.

The calibrated hyper-parameter values for the prior distributions are given in Table~\ref{tab:prior_par}. Details of the grid over which the search was conducted are available in the supplementary materials. In many methods these values are the same in setting 1 and setting 2 (TITE-CRM, TITE-CRM2, POMM, TITE-mTPI2, R-mTPI2) and for the other methods they are relatively similar. There are some differences between these values and those in the original proposals, discussed further in Section~\ref{sec:results}

\section{Simulation Studies} \label{sec:simulations}
In this section we detail the comparative simulation studies undertaken. We first describe the data generation used in the simulation studies in Section \ref{sec:data_generation}, then present and analyse the results of those studies in Section \ref{sec:results}.
\subsection{Data Generation} \label{sec:data_generation}
Since the methods require different levels of information of patient response, the mechanism of generation of patient responses is not immediately obvious. Here we describe the process of generating data in a generic way. For notational simplicity, we describe the generation for a single dose and so have omitted any index referring to dose.

Data are generated for each patient in the following way. Each patient $i$ has a latent toxicity variable $z_i$ drawn from a Uniform(0,1) distribution. This variable determines the outcome of patient $i$ on all cycles and all doses. 

It is assumed that there is a constant decrease in P(DLT) across cycles, this value is taken to be 1/3. Extending the notation of defining $p_s$ as the true $P(DLT)$ in cycles up to and including cycle $s$, we define $p$ as the true $P(DLT)$ in the entire follow up period. We obtain the following for a follow up period of 3 cycles:
\[
p=p_3=p_1 + (1-p_1)\frac{p_1}{3} + (1-p_1)(1-\frac{p_1}{3})\frac{p_1}{9}.
\]
For example, given a $p_1=0.3$, we would have $p_3=0.391$.

To generate the binary variable $Y_{i,s}$, which equals 1 if patient $i$ observes a DLT response in cycle $s$ and 0 otherwise, the simple indicator is used $Y_{i,s}= \mathds{I}[p_{s-1}<1-z_i<p_s]$ where $p_0=0$ and $Y_{i,1}$ is always defined, with $Y_{i,s}$ only defined for $s>1$ if $Y_{i,s-1}=0$.

Both the POMM and the nTTP approaches need more detailed patient responses than the binary variable $Y_{i,s}$, and so we must also generate the grades and types of toxicities. The nTTP method specifies three type of toxicity (renal, haematological, neurological) of grades 0-4. Patients may observe toxicities of different types, and the maximum grade is used in the POMM approach. As is  standard in such studies, we classify a toxicity of grade 3 or above as a DLT. 

We first illustrate how the maximum observed grades are calculated for the first cycle. First, define the probability of a grade $g$ toxicity being the maximum observed grade in cycle 1 as $p^{Gg}_1$. Then let 
\begin{align*}
p^{G4}_1&=p^{G3}_1=p_1/2 \\
p^{G2}_1&=(1-p_1)\hspace{1pt}\mathds{I}[p_1 > \frac{1}{2}] + p_1 \hspace{1pt} \mathds{I}[p_1 \leq \frac{1}{2}] \\
p^{G1}_1&=(1-2p_1)\hspace{1pt} \mathds{I}[\frac{2}{5}< p_1 <\frac{1}{2}] + \frac{p_1}{2} \hspace{1pt}\mathds{I}[p_1 \leq \frac{2}{5}] \\
p^{G0}_1&=(1-\frac{5p_1}{2})\hspace{1pt} \mathds{I}[p_1 \leq \frac{2}{5}]
\end{align*}
In the same way as $Y_{i,1}$ is calculated, the observed grade $Y^{Gg}_{i,1}$ is determined by the latent variable $z_i$ in the following way:
\begin{align*}
  Y^{Gg}_{i,1}&=\mathds{I}[\sum^{4}_{k=g+1} p^{Gk}_1 < 1-z_i < \sum^{4}_{k=g}p^{Gg}_1] \mbox{\hspace{10pt} for $g < 4$, and}\\ 
  Y^{G4}_{i,1}&=\mathds{I}[ 1-z_i < p^{G4}_1].
  \end{align*}
To illustrate this, we provide an two examples, one dose where the probability of DLT in cycle 1 is exactly on target at 0.3, and a second dose where the probability of DLT in cycle 1 is 0.6. 

\textit{Example 1:}
\begin{align*}
p^{G4}_1&=p^{G3}_1=0.15 \\
p^{G2}_1&=0.3 \\
p^{G1}_1&=0.15 \\
p^{G0}_1&=0.25
\end{align*}
\textit{Example 2:}
\begin{align*}
p^{G4}_1&=p^{G3}_1=0.3 \\
p^{G2}_1&=0.4 \\
p^{G1}_1&=0 \\
p^{G0}_1&=0
\end{align*}
Next, we describe how we calculate the combination of grade and type of observed toxicity. For the five grades and three types, there are therefore 125 combinations of type and grade observations. These are partitioned into sets defined by the maximum grade, with the probability that the observation is in a given set being the previously defined $p^{Gg}_1$. Within each set, each combination has equal probability. $nTTP$ values are then calculated according to the weights given by Yin et al. \cite{Yin2017}.

For subsequent cycles, the same approach is taken, with all $p^{Gg}_s$ values scaled accordingly, so that $p^{Gg}_2=(1-p_1)p^{Gg}_1$ and $p^{Gg}_3=(1-p_2)p^{Gg}_1$. This means that for patients with $1-z_i<p_s$, no responses are defined for cycles after cycle $s$, since the patient has left the study. In this framework of generating data, the probability of each outcome is defined by the value of $p_1$.
\subsection{Results} \label{sec:results}
5000 simulations are conducted for each approach across a range of scenarios. A full study of 17 scenarios is undertaken in order to explore the behaviour of the different methods in a wide range of scenarios. We focus on the results of four of these scenarios here, with full specification and results of all scenarios given in the online supplementary materials. The target, $\tau$, of the P(DLT) across the three cycles is 0.391, equating to P(DLT) in cycle 1 of 0.3 in our data generation.

\begin{figure}[h!]
\centering
\subfigure{\includegraphics[width=0.46\textwidth]{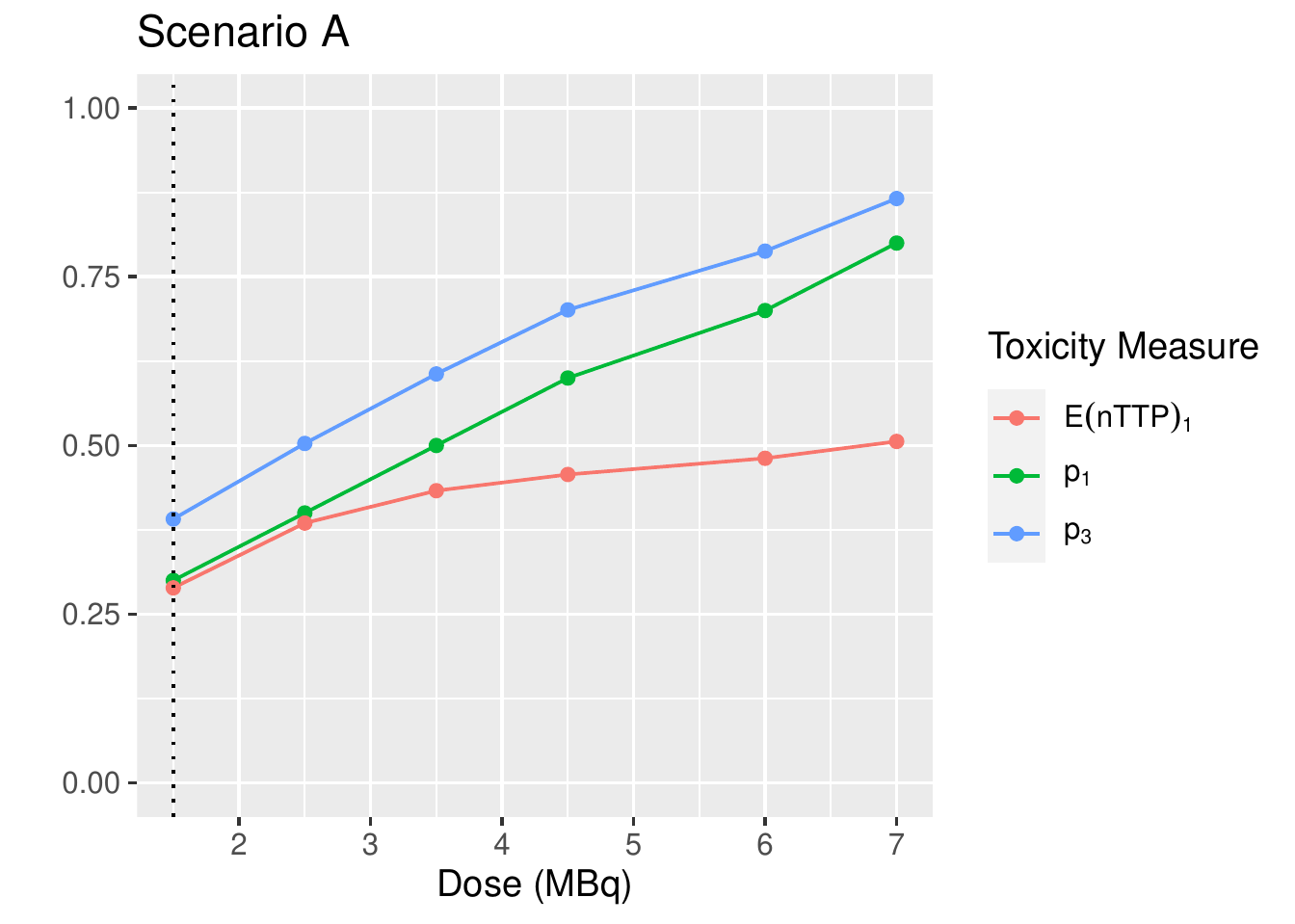}}
\hspace{0.05\textwidth}
\subfigure{\includegraphics[width=0.46\textwidth]{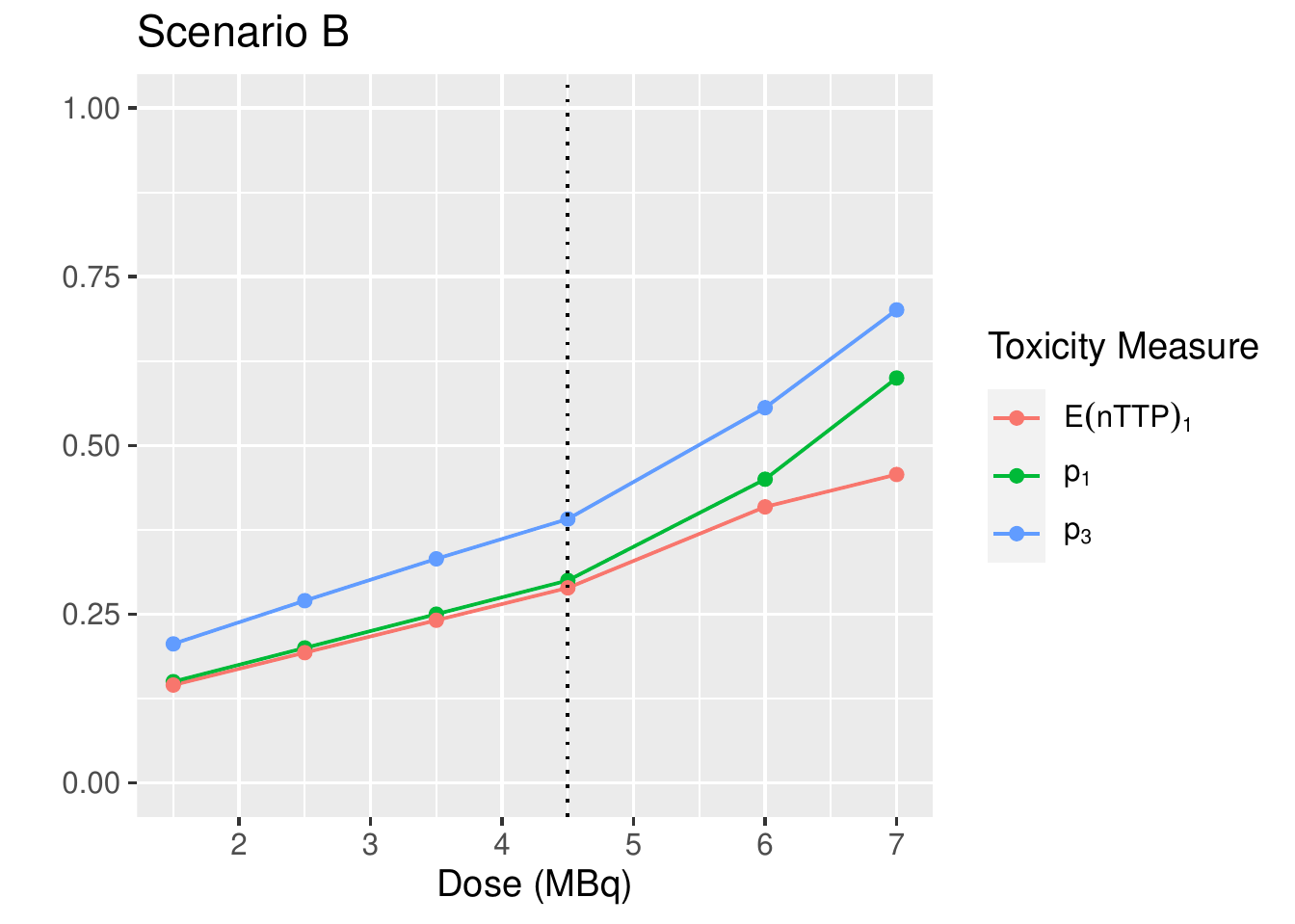}}
\subfigure{\includegraphics[width=0.46\textwidth]{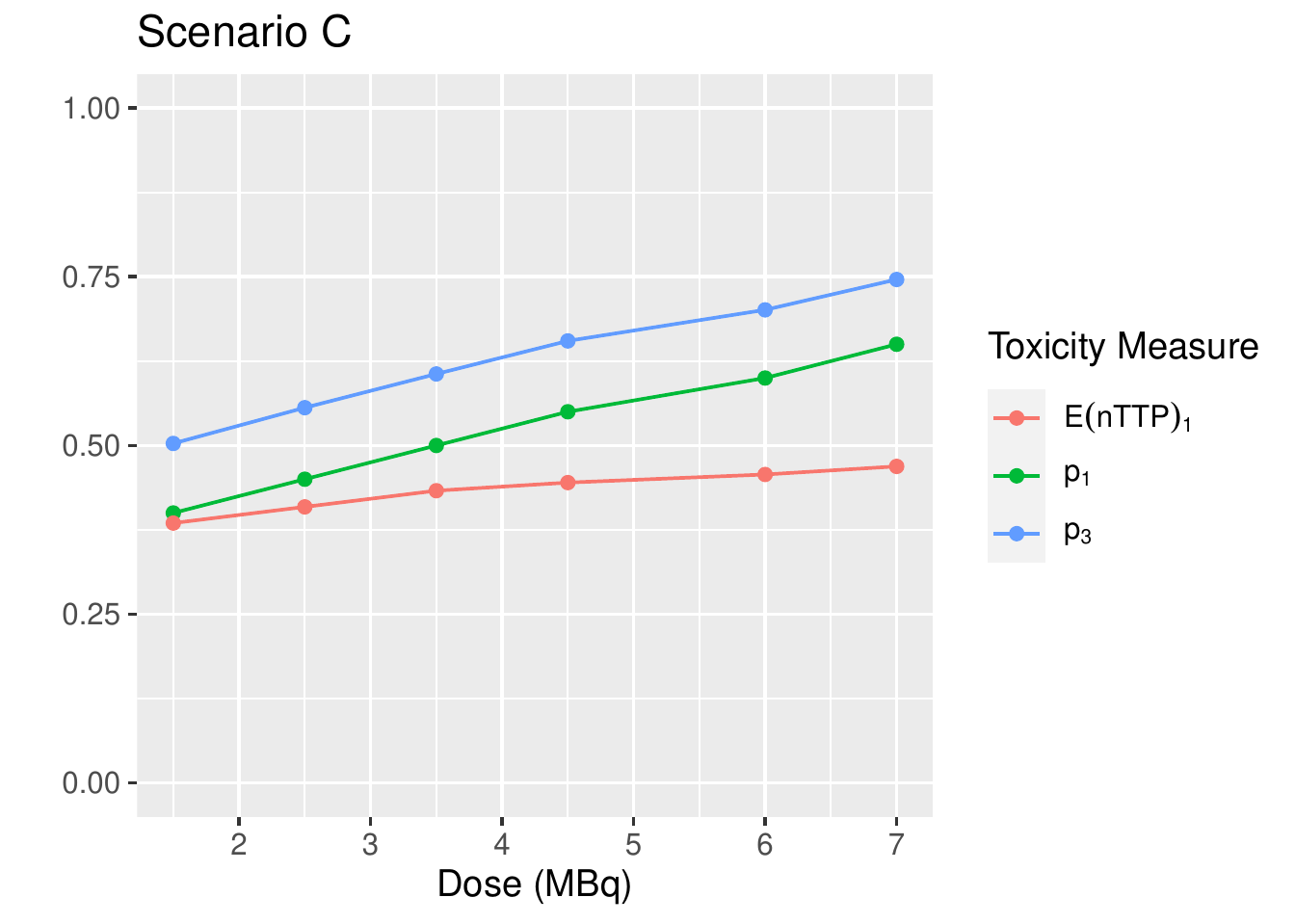}}
\hspace{0.05\textwidth}
\subfigure{\includegraphics[width=0.46\textwidth]{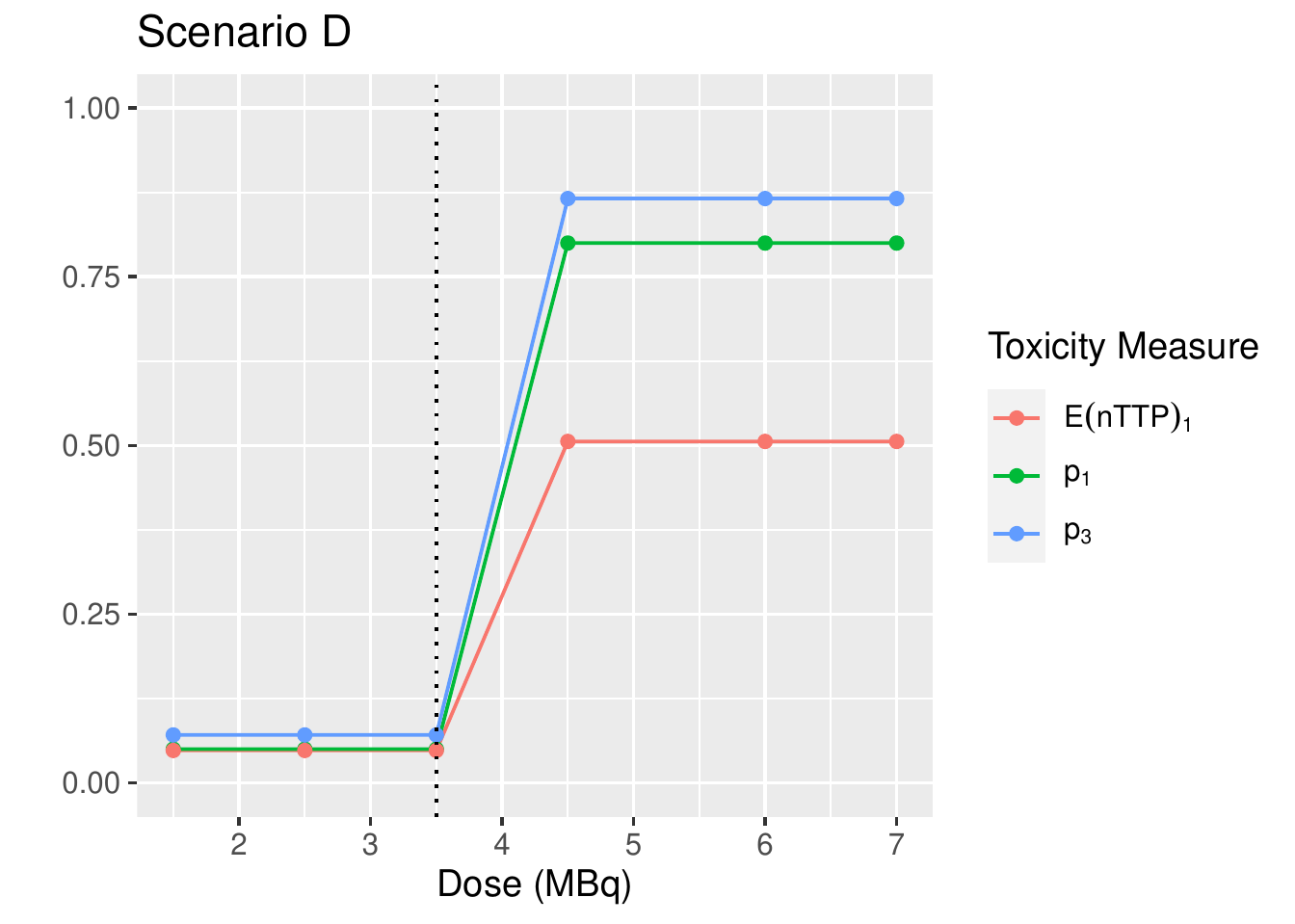}}
\caption{Specification of the four scenarios, with MTD highlighted in with a dotted line for scenarios A, B and D. $E(nTTP)_1$ is the expected nTTP value for cycle 1. } \label{fig:scenarios}
\end{figure}

The four scenarios we look at in depth are specified in Figure~\ref{fig:scenarios}, with the MTD in scenarios A, B and D highlighted by the dotted line. In scenario A, the lowest investigated dose is the MTD, with $p_1$ linearly increasing with dose level. In scenario B, the fourth dose level is the MTD, with a non-linear increase in $p_1$ by dose level. In both of these scenarios, the MTD has the exact target $p_1$. In scenario C, all doses are unsafe. In scenario D, the lowest three doses have $p_1=0.05$ and the highest three doses have $p_1=0.8$, hence the third dose level is the MTD. Although $p_1$ is clearly well below the target of 0.3, the fourth dose is very unsafe and so the third dose level is by definition the maximum dose that is on target or below.

We use a number of metrics to compare the performance of the designs. The first is the proportion of correct selections (PCS), defined as the proportion of simulations that make the correct choice, be that recommending the true MTD or stopping for safety when all doses are unsafe. Note that in the supplementary materials, where we also consider scenarios in which all doses have $P(DLT)$ below the target $\tau$, we define a correct outcome in this case as stopping the trial for stopping rule 3, highest dose deemed very safe.  We compare this PCS value to an empirical optimal benchmark \citep{oquigley2002}, whereby each individual patient's latent toxicity variable $z_i$ determines that patient's response in all cycles on all dose levels. This is evaluated for all patients, as we know the response of any patient at any possible dose level and the dose with the mean response closest to the target (either nTTP or P(DLT) across the three cycles) is chosen as the recommended dose. The benchmark level is then the number of simulations that correctly identify the MTD with this full information on all patients and all doses. It is important to note that the way the simulations are conducted means that it is the same sequence of patients with the same latent toxicity variables in the simulations for all methods and the benchmark comparator. The only difference is that the benchmark always uses the maximum number of patients, whereas the other methods have the options to stop before this maximum is reached. The purpose of this benchmark is to give and indication of the difficulty of the scenario, with more difficult scenarios exhibiting a lower percentage of PCS.

We use two measures of the size of the trial: the total number of patients and the total length of the trial in weeks until all recruited patients have finished their follow up time or experienced DLTs. It is desirable to have a shorter trial with fewer patients. We consider the allocations of patients, focussing on the number of patients assigned to the MTD and to unsafe doses. The reason for stopping the trial is also of interest, especially in setting 2 where there are many stopping rules implemented.

\begin{figure}[h!]
\centering
\subfigure[Scenario A]{\includegraphics[width=0.46\textwidth]{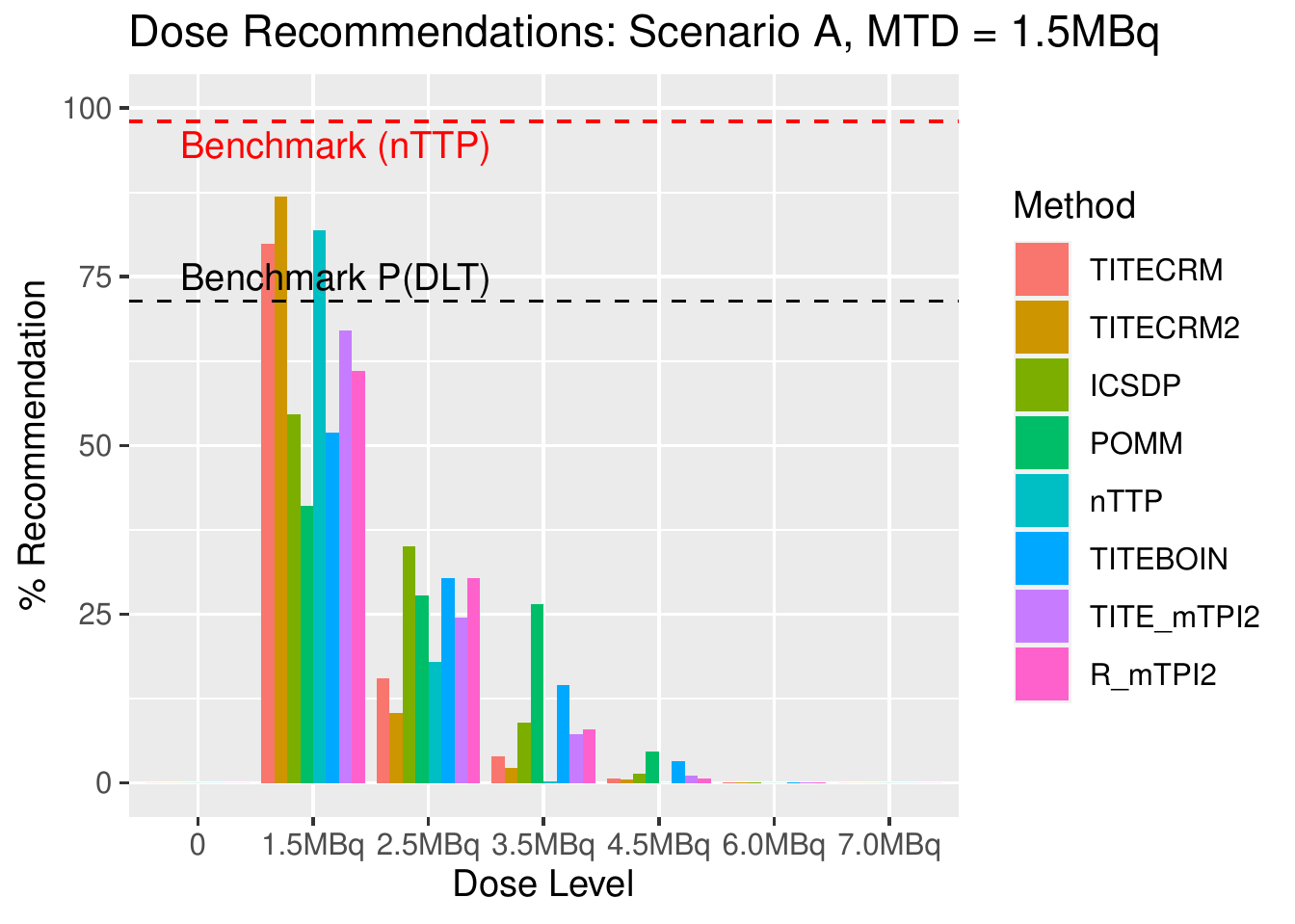}\label{fig:sA_stage1_rec}}
\hspace{0.05\textwidth}
\subfigure[Scenario B]{\includegraphics[width=0.46\textwidth]{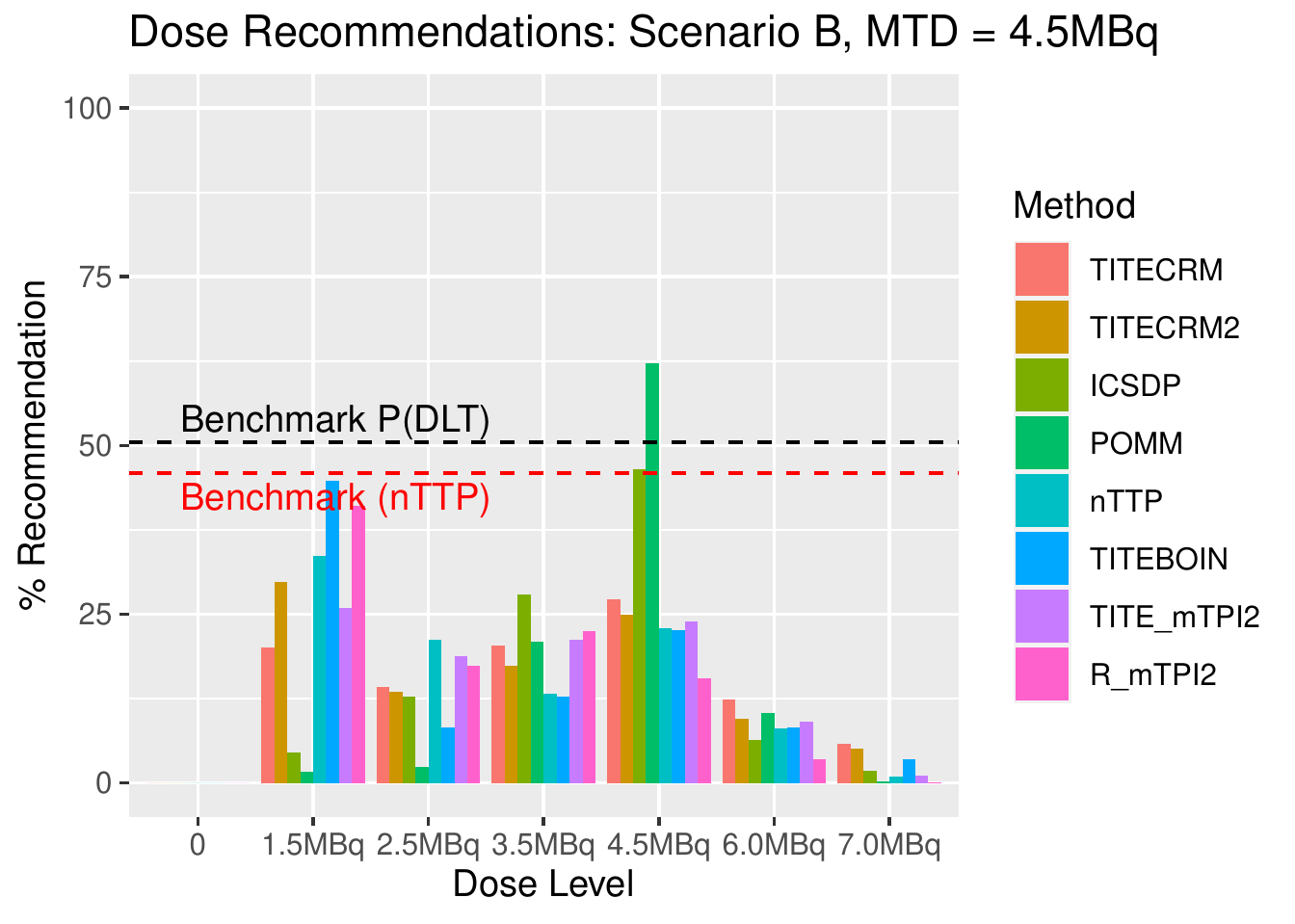}\label{fig:sB_stage1_rec}}
\subfigure[Scenario C]{\includegraphics[width=0.46\textwidth]{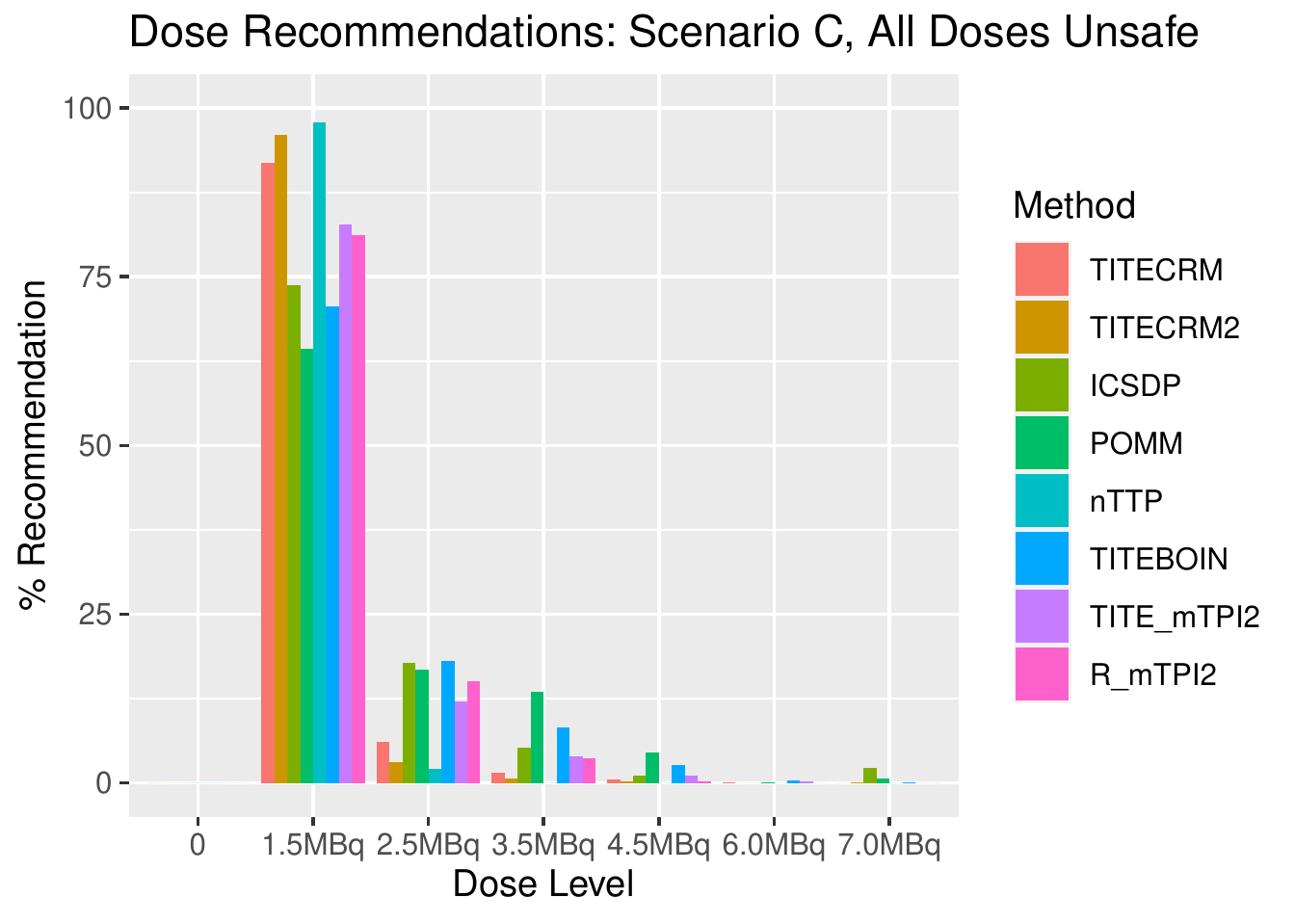}\label{fig:sC_stage1_rec}}
\hspace{0.05\textwidth}
\subfigure[Scenario D] {\includegraphics[width=0.46\textwidth]{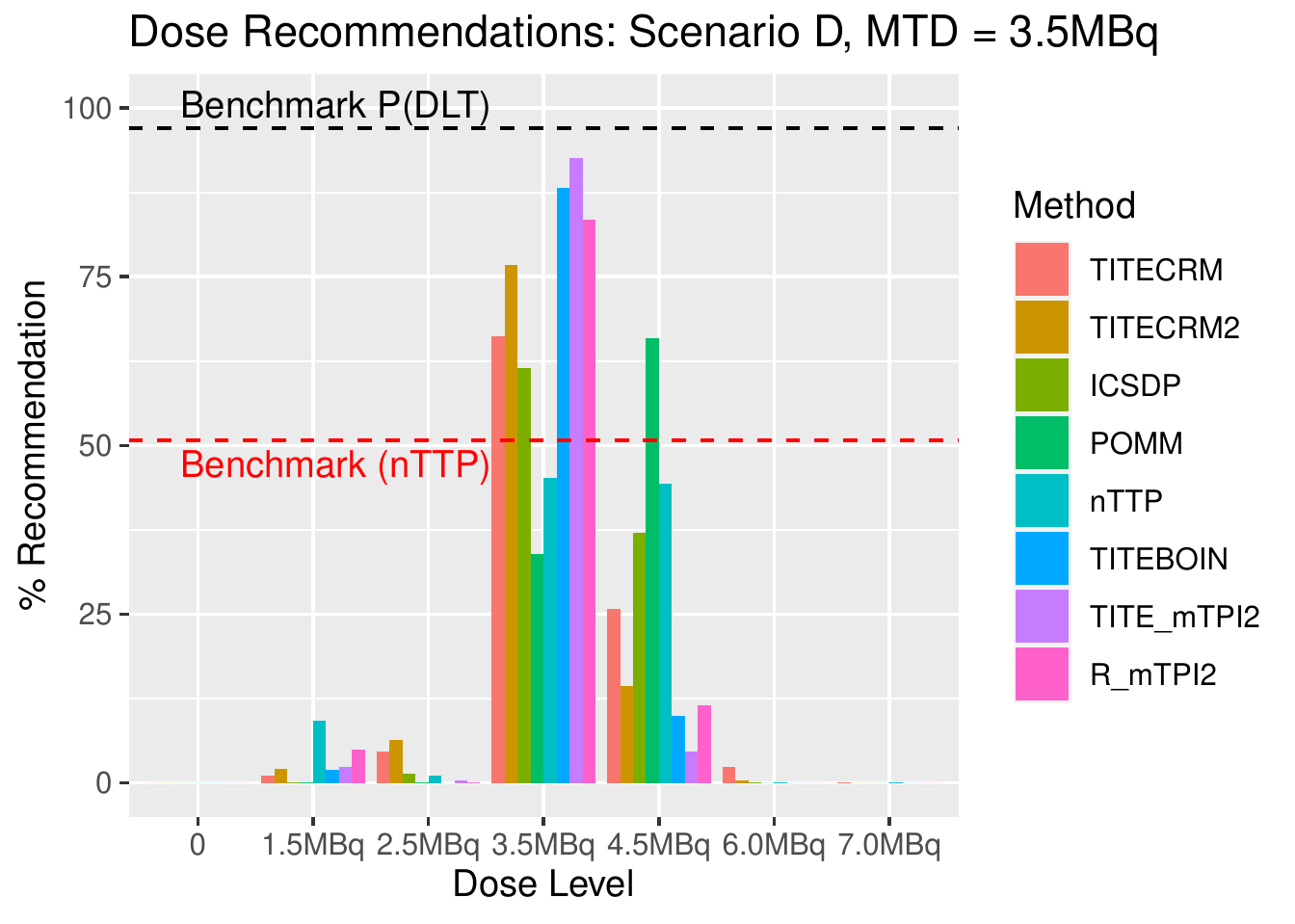}\label{fig:sD_stage1_rec}}
\caption{Setting 1 dose recommendations expressed as a percentage of simulations which recommend the given dose level. 0 indicates no dose is recommended, which is not applicable in setting 1.} \label{fig:stage1_recs}
\end{figure}

\begin{figure}[h!]
\centering
\subfigure[Scenario A]{\includegraphics[width=0.46\textwidth]{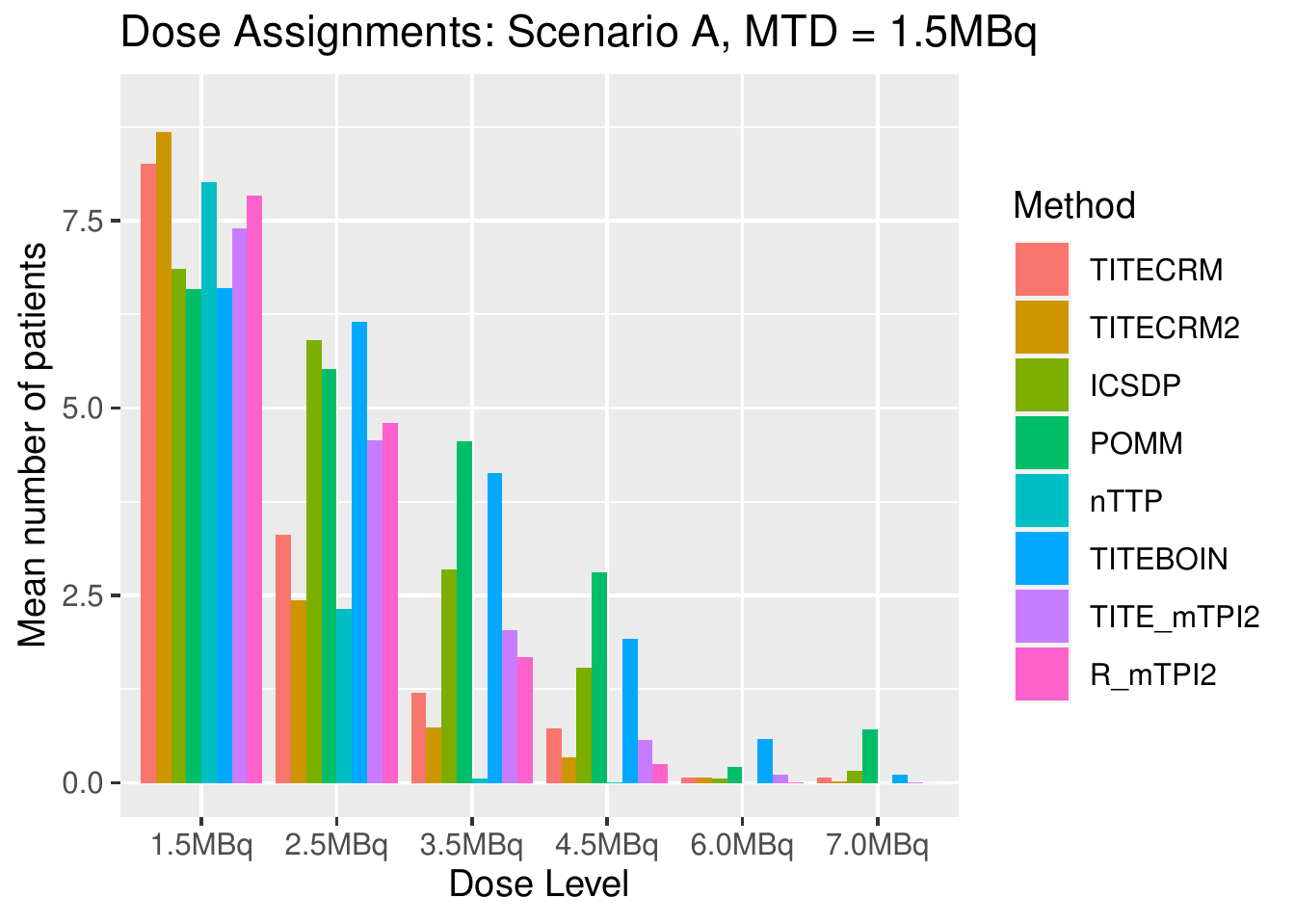}\label{fig:sA_stage1_ass}}
\hspace{0.05\textwidth}
\subfigure[Scenario B]{\includegraphics[width=0.46\textwidth]{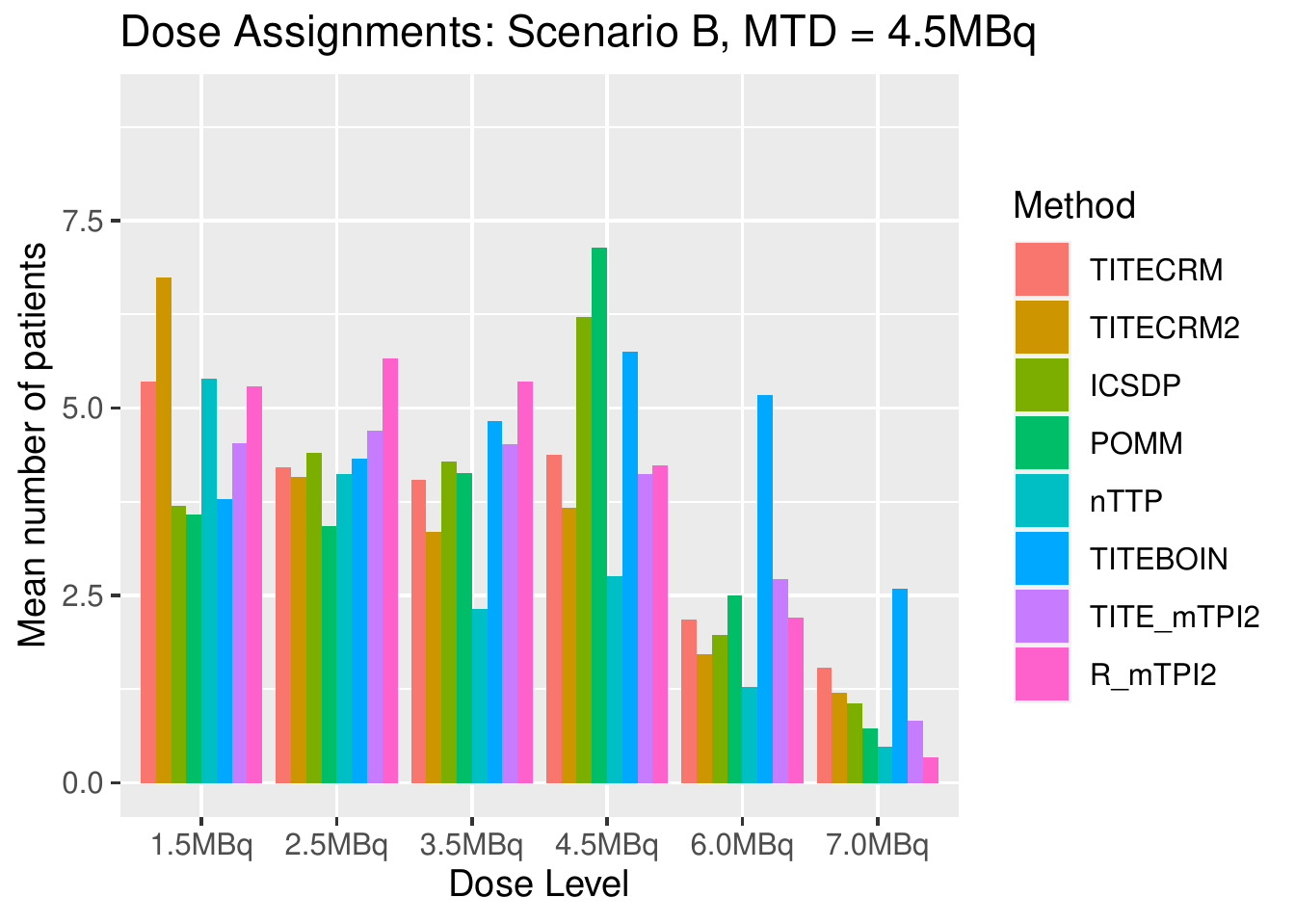}\label{fig:sB_stage1_ass}}
\subfigure[Scenario C]{\includegraphics[width=0.46\textwidth]{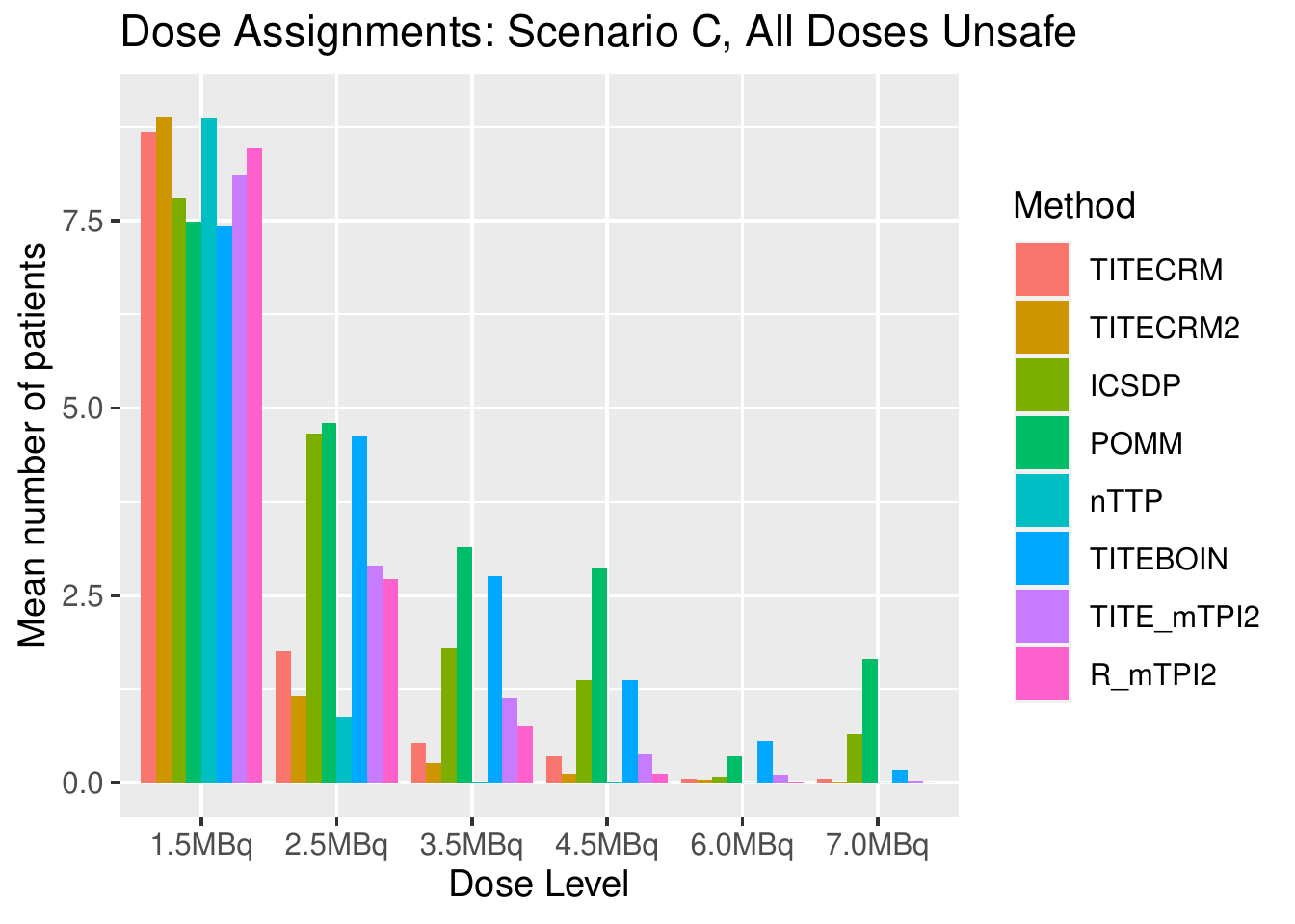}\label{fig:sC_stage1_ass}}
\hspace{0.05\textwidth}
\subfigure[Scenario D] {\includegraphics[width=0.46\textwidth]{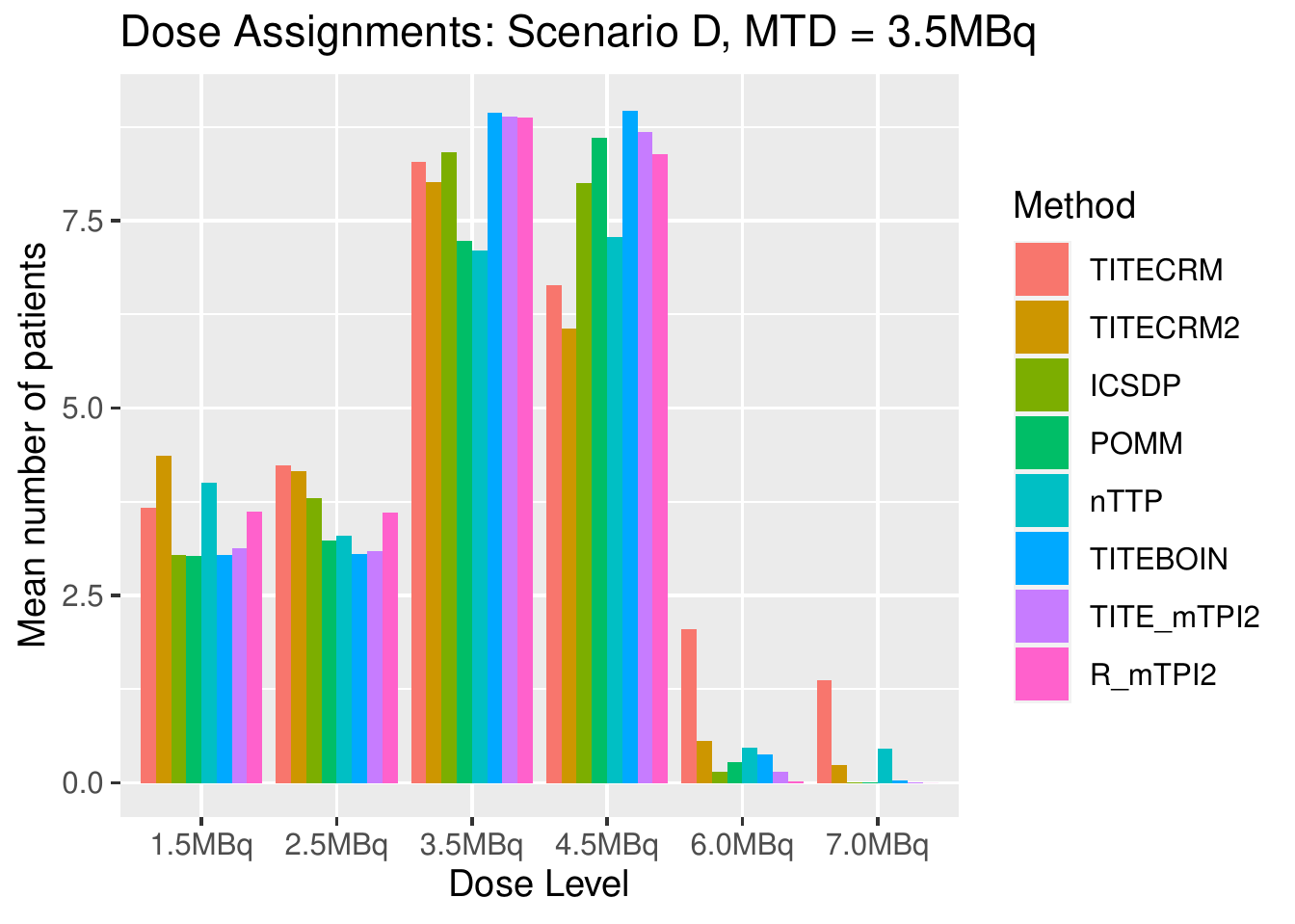}\label{fig:sD_stage1_ass}}
\caption{Setting 1 dose assignments expressed as an average number of patients over simulations which are assigned the given dose level.} \label{fig:stage1_ass}
\end{figure}

\begin{figure}[h!]
\centering
\subfigure[Scenario A]{\includegraphics[width=0.46\textwidth]{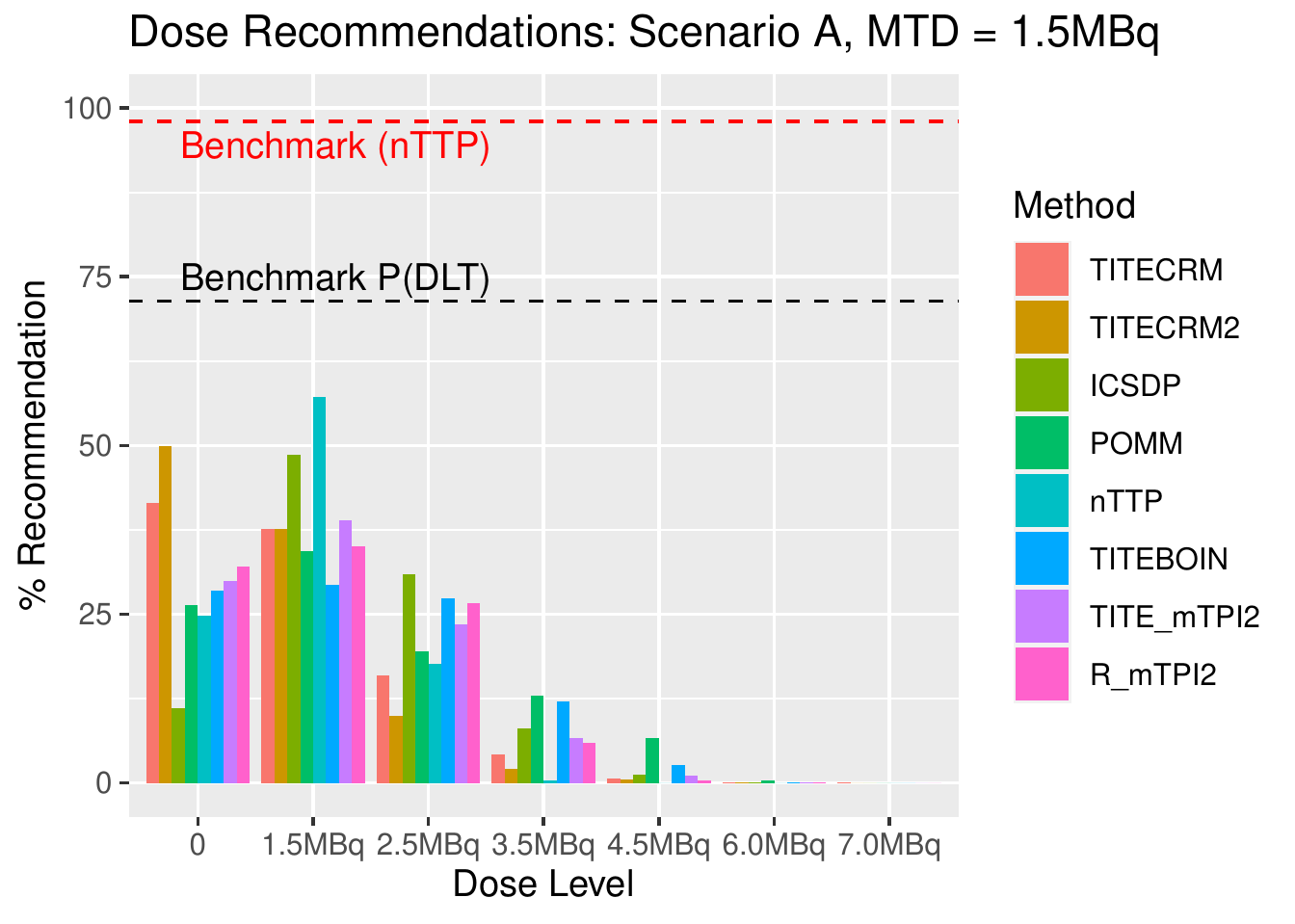}\label{fig:sA_stage2_rec}}
\hspace{0.05\textwidth}
\subfigure[Scenario B]{\includegraphics[width=0.46\textwidth]{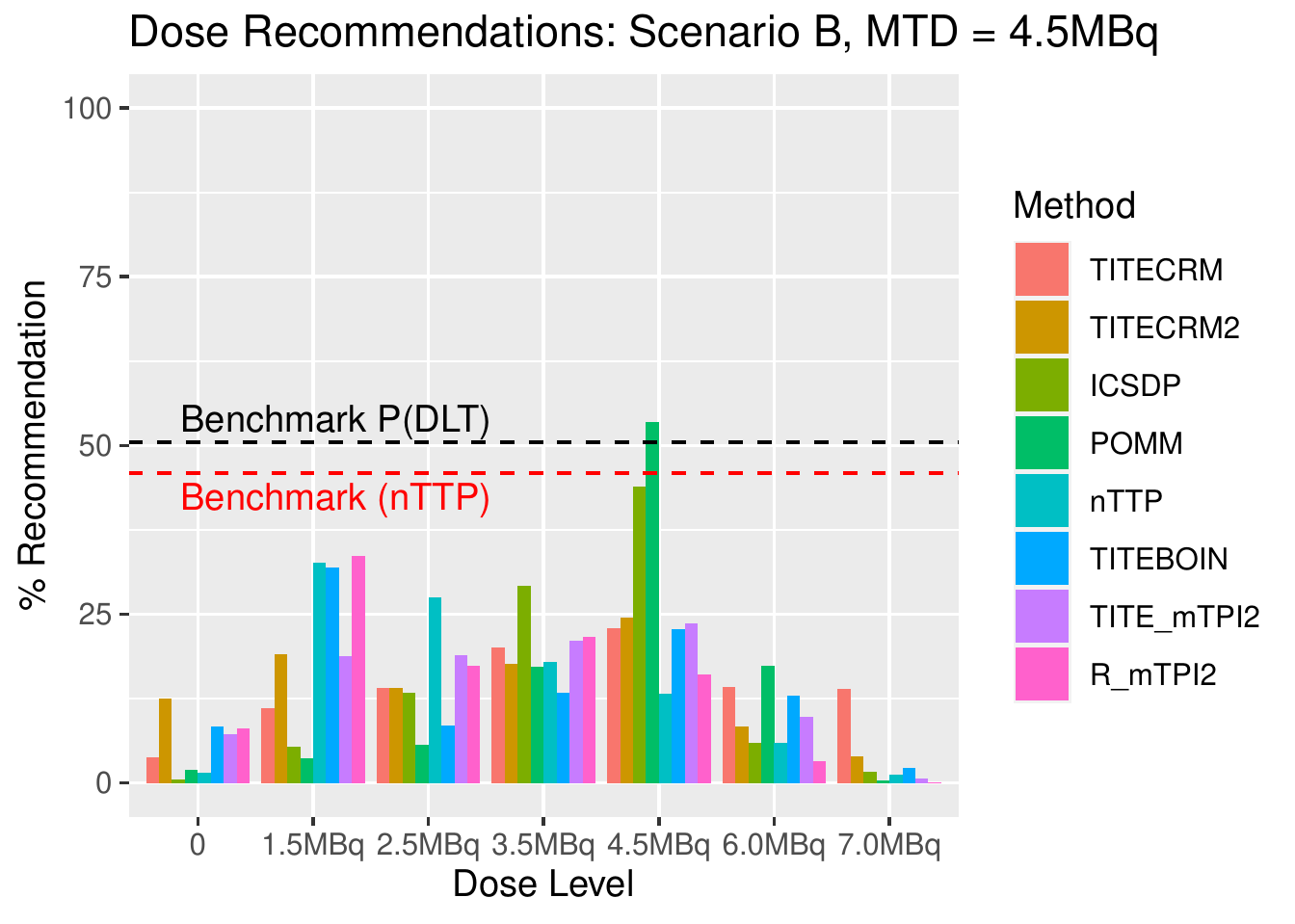}\label{fig:sB_stage2_rec}}
\subfigure[Scenario C]{\includegraphics[width=0.46\textwidth]{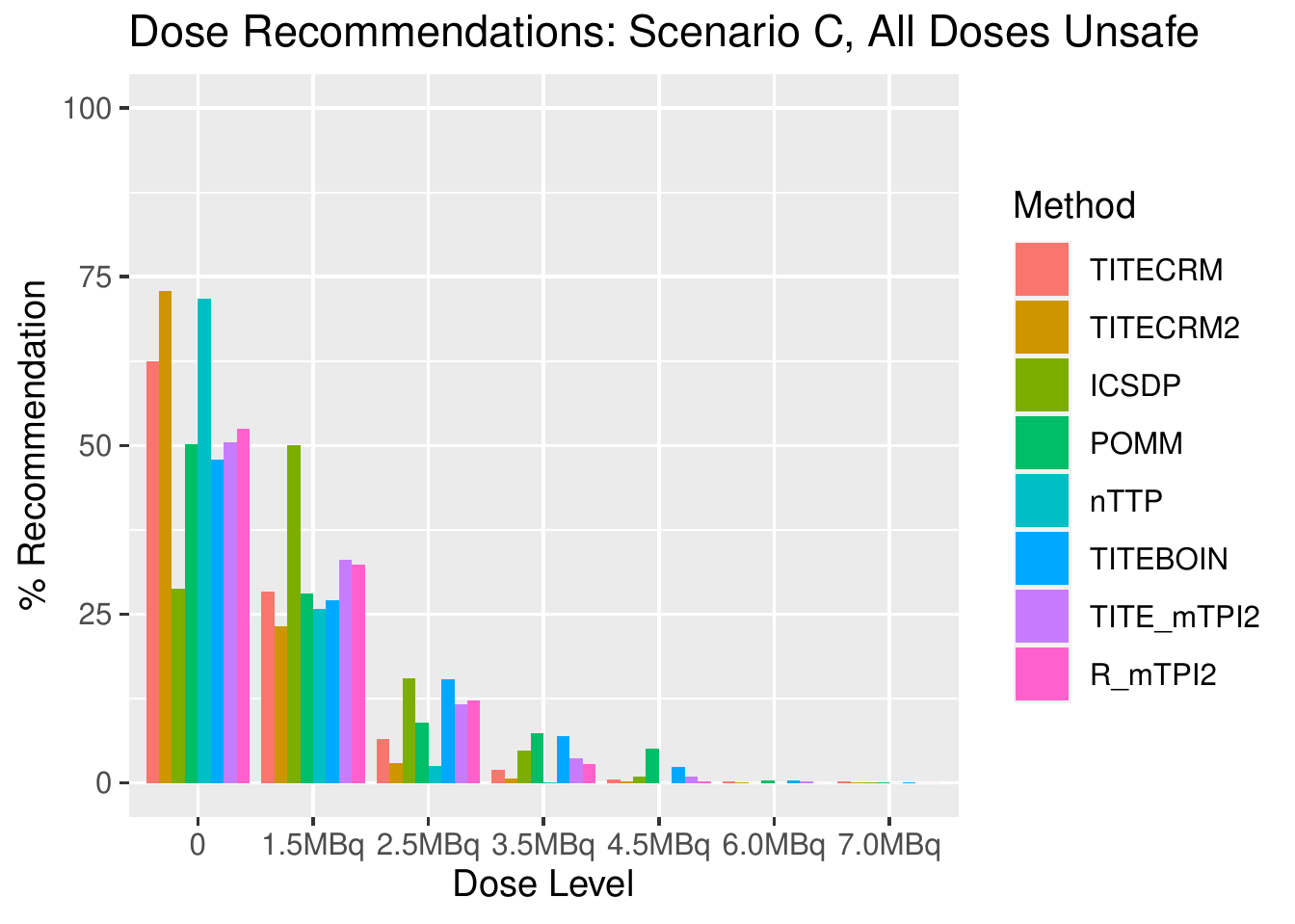}\label{fig:sC_stage2_rec}}
\hspace{0.05\textwidth}
\subfigure[Scenario D] {\includegraphics[width=0.46\textwidth]{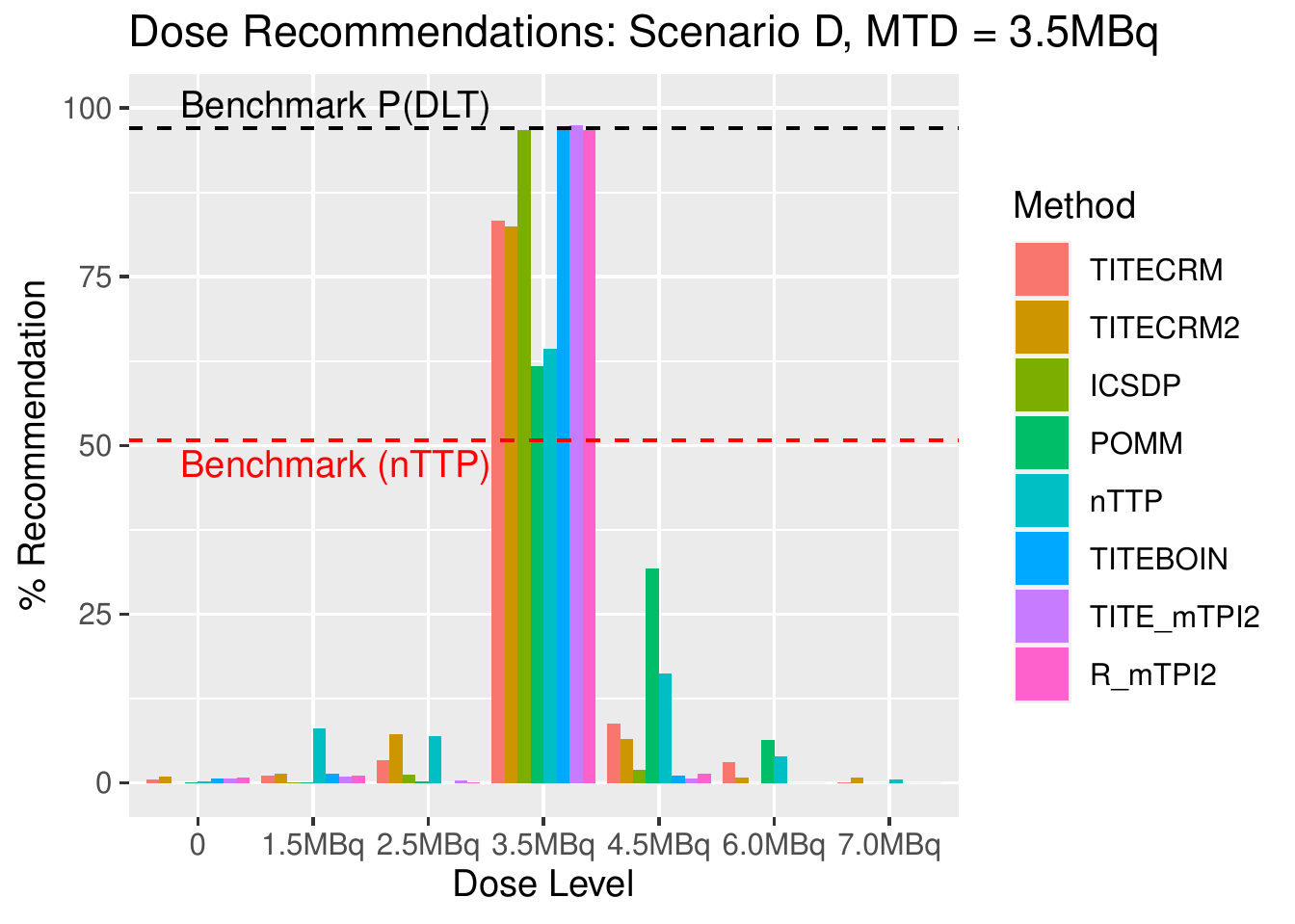}\label{fig:sD_stage2_rec}}
\caption{Setting 2 dose recommendations expressed as a percentage of simulations which recommend the given dose level. 0 indicates no dose is recommended.} \label{fig:stage2_recs}
\end{figure}

\begin{figure}[h!]
\centering
\subfigure[Scenario A]{\includegraphics[width=0.46\textwidth]{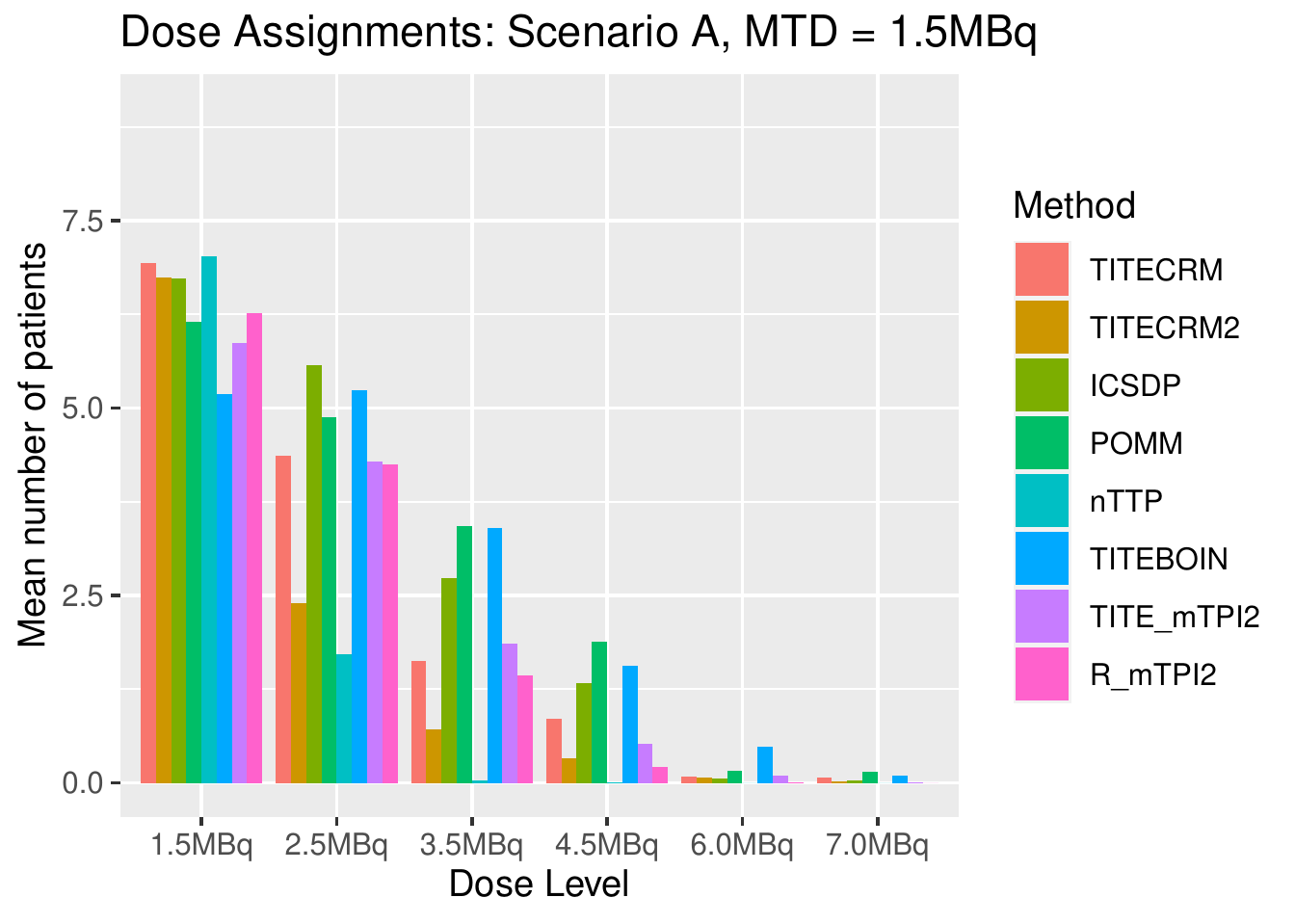}\label{fig:sA_stage2_ass}}
\hspace{0.05\textwidth}
\subfigure[Scenario B]{\includegraphics[width=0.46\textwidth]{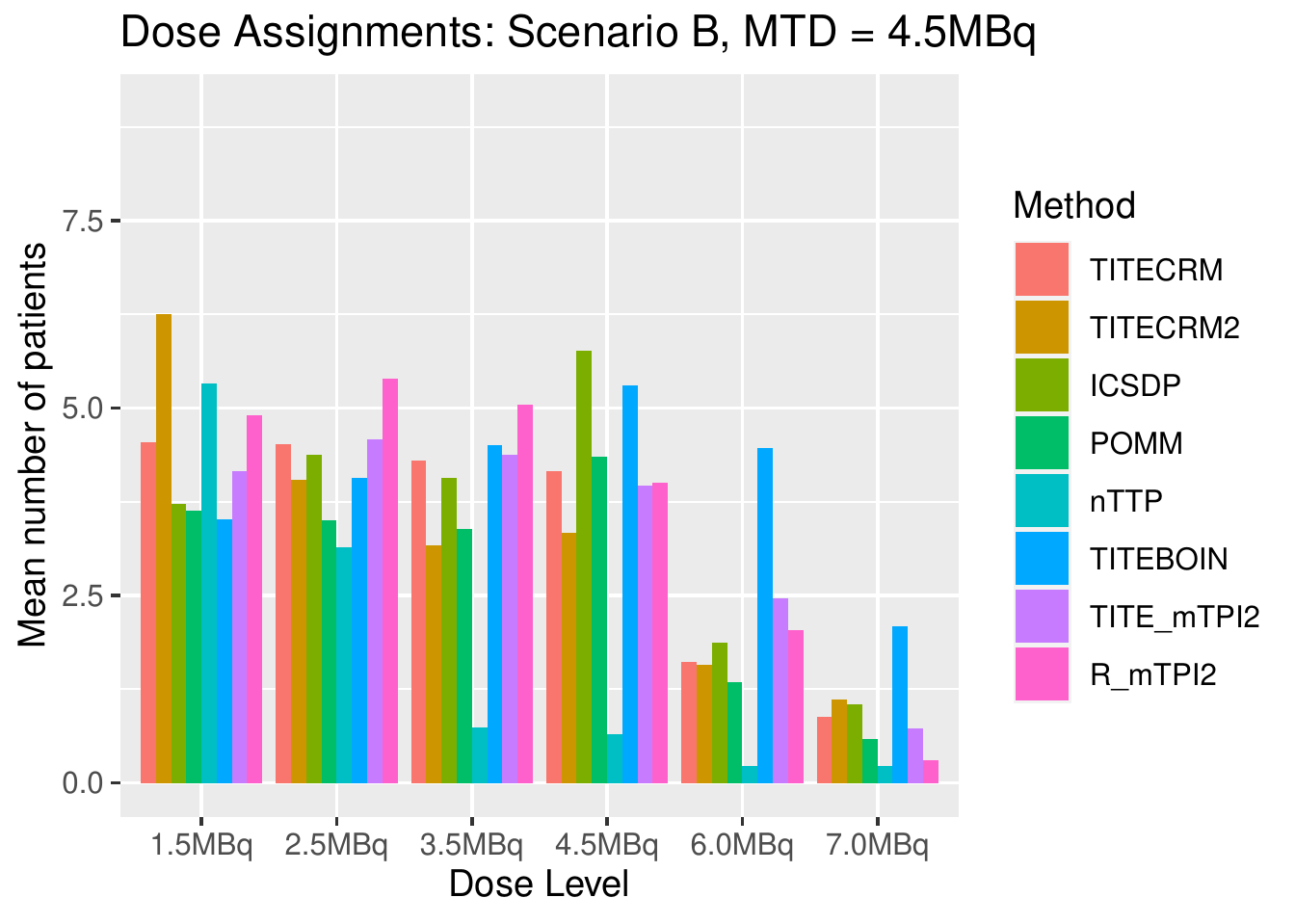}\label{fig:sB_stage2_ass}}
\subfigure[Scenario C]{\includegraphics[width=0.46\textwidth]{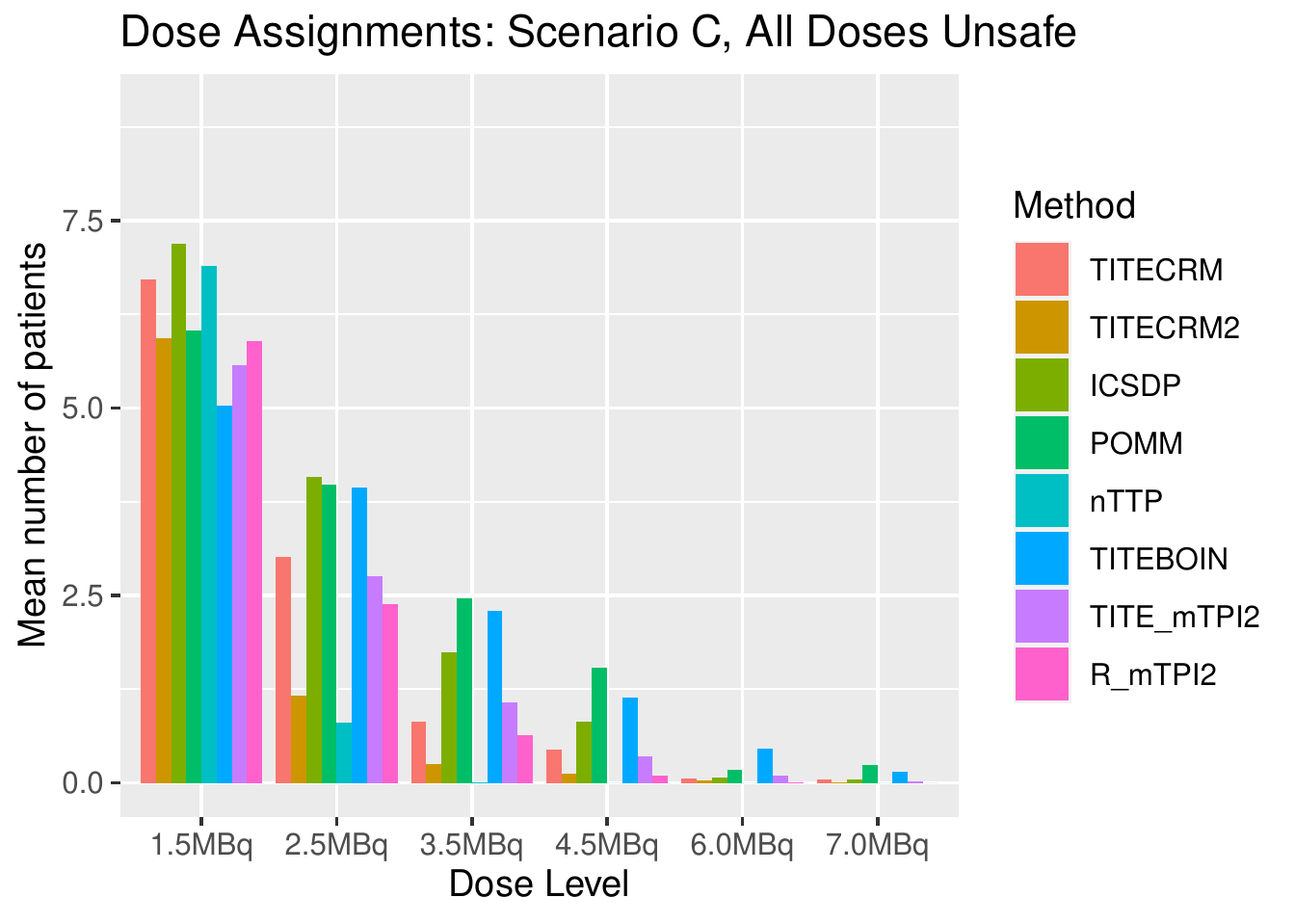}\label{fig:sC_stage2_ass}}
\hspace{0.05\textwidth}
\subfigure[Scenario D] {\includegraphics[width=0.46\textwidth]{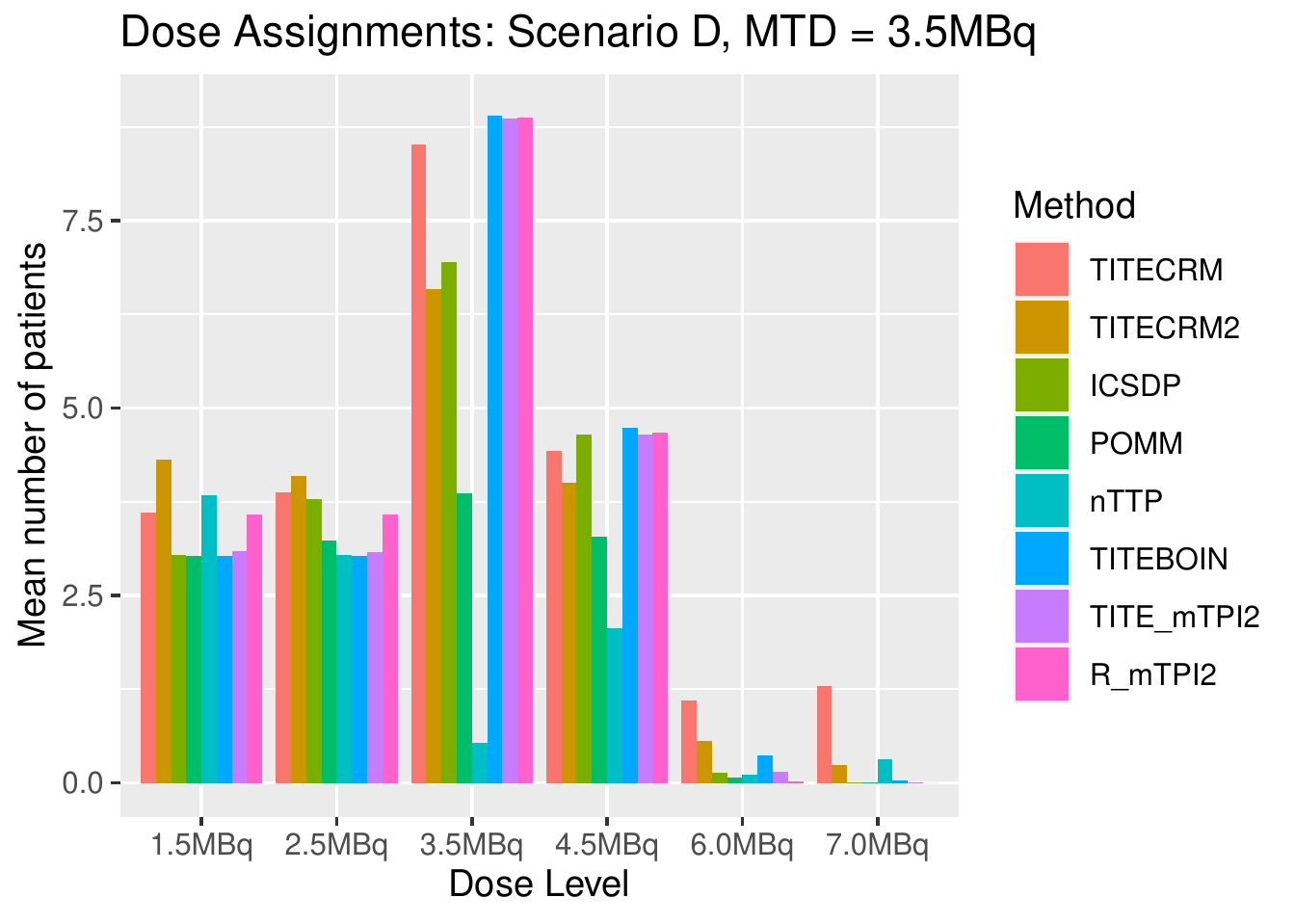}\label{fig:sD_stage2_ass}}
\caption{Setting 2 dose assignments expressed as an average number of patients over simulations which are assigned the given dose level.} \label{fig:stage2_ass}
\end{figure}

In scenario A, where the lowest dose is the MTD, the best performing design in setting 1 in terms of PCS is the TITE-CRM2 (87\%), closely followed by the nTTP (82\%) and TITE-CRM (80\%), these are also the designs with the highest allocation to the the true MTD (a mean of 9, 8 and 8 respectively). These designs even outperform the benchmark, a phenomenon possible due to the sufficient information stopping rule. The worst performing design in this case is the POMM (41\%), where a large number of patients are assigned to unsafe doses, on average 14. However in scenario B, where the fourth dose level is the true MTD, the POMM is the best performing design in setting 1 with a PCS of 62\%. This could also potentially be driven by the prior pseudo data that in this case closely matches the true scenario for the lowest four out of the six investigated doses. The ICSDP is the next best performing with a PCS of 47\%, with the other designs all below 30\%. In this scenario we also see that both variants of the mTPI2 design have a much longer trial duration, nearly twice as long as the model-based designs.

In scenario C, where all doses are unsafe, setting 1 does not allow for stopping for safety (this is left for setting 2), therefore there is no measure of PCS. However we can still note that the POMM has the largest mean sample size of 20 patients and mean duration of 52 weeks. We can also see in Figure~\ref{fig:stage1_recs}, that due to the lack of stopping rules in setting 1, the shape of the graphs for scenario A and C are very similar. In scenario D, where the third dose level is the MTD, the model-assisted methods show superior performance in terms of PCS, ranging from 83-93\%. There is no noticeable difference in sample size between these designs and the model-based ones, but there is still a large increase in the mean trial duration for the mTPI2 based designs. Figure~\ref{fig:stage1_ass} shows that for most methods, there is a similar level of assignment to the third and fourth doses, despite the vast differences in underlying toxicity between the two doses. 

In setting 1, minimal stopping rules are implemented in order to investigate the behaviour of the designs by themselves. In setting 2 however, the rules are more reflective of a true dose-finding trial, with stopping rules for safety and precision. In scenario A, this is especially noticeable in that it introduces the extra possibility of wrongly stopping the trial because the lowest dose is deemed unsafe. Comparing Figure~\ref{fig:sA_stage1_rec} to Figure~\ref{fig:sA_stage2_rec} it is clear that for all methods, a large proportion of simulations that in setting 1 correctly recommended the lowest dose as the MTD, now stop early for safety. This is especially prevalent for the TITE-CRM2 method, the best performing in setting 1, where 50\% of simulations are stopped for this reason. The best performing method in scenario A in setting 2 is the nTTP, although a PCS of 57\% in a scenario where the lowest dose is the MTD is by no means an outstanding performance. 

Scenario B shows less of a contrast in setting 2 to setting 1, with POMM and ICSDP giving the best, and similar performances. In scenario C, the safety stopping rules are implemented more effectively in some methods than others. Both the TITE-CRM2 and nTTP stop for safety in around 72-73\% of simulations, whereas the ICSDP only stops for safety in 29\% of simulations, with 50\% of simulations recommending the lowest dose as the MTD. There is an average of 14 patients, nearly 5 cohorts, an unacceptable level when all doses are unsafe. This is most likely driven by the prior for this method, as there are 5 pseudo-patients on the lowest dose, providing evidence that although chosen by the calibration procedure, such a prior may be too strong to use practically.  The POMM also has an average of 14 patients in this scenario, although stops in 50\% of simulations. The dose assignment for this method is more unsafe than the ICSDP, indicated in Figure~\ref{fig:sC_stage2_ass}, by the high levels of assignment to higher dose levels. The TITEBOIN design also sees high levels of assignment to higher doses.

In scenario D, the PCS is improved in setting 2 over setting 1, with four methods (ICSDP, TITEBOIN, TITE-mTPI2 and R-mTPI2) achieving the benchmark level and nTTP even exceeds this. This is in part due to the hard safety rule that eliminates unsafe doses, and in part to the precision stopping rule. The dose assignment shows this is the case, comparing Figure~\ref{fig:sD_stage2_ass} to Figure~\ref{fig:sD_stage1_ass} clearly shows the reduction in assignment to the fourth dose. 

In both settings, the mTPI2 based methods clearly have a longer than ideal trial length and hence would not be recommended for use. The TITEBOIN also has a larger patient numbers in most scenarios. It is important to note that the relationship between trial duration and the total trial size for the methods varies due to the rules implemented in the different methods. The model-assisted approaches' PCS are also less than the model-based approaches in most scenarios in setting 1, although this is improved somewhat in setting 2. Of the model-based approaches, although the nTTP provides the shortest trial duration, the PCS is much lower than the other designs when the true MTD is in the higher doses in setting 2, and so also not recommended. This is mainly due to premature stopping for precision, a consequence of the fact that this rule is designed for binary rather than continuous endpoints. In terms of overall performance across the full range of scenarios, the ICSDP is the front-runner, despite its poor performance when all doses are unsafe in setting 2. The next best performing method is TITE-CRM2, which gives a good yet balanced performance across scenarios in both settings. The TITE-CRM2 also has a slightly shorter average trial duration, requiring fewer patients overall and fewer patients treated on unsafe doses. However we also note that the process of data generation follows more closely the ICSDP assumptions than the TITE-CRM2, and so this comparison may  be slightly different under the different assumptions.
\section{Discussion} \label{sec:discussion}
In this paper, we conducted a simulation study to compare the leading methods for dose-finding trials incorporating later onset toxicities in a variety of scenarios. The purpose of such a comparison was to evaluate the performance of the different methods in generic settings where their individuals assumptions may not hold, in order to highlight any key differences.

The values of the hyper-parameters for the prior distributions in each method were calculated using a calibration procedure. They were calibrated over a small number of clinically plausible scenarios, but still ranging in the position of the MTD in the dosing sequence. This ensured that the different methods all had the same opportunity to achieve their potential in a large range of scenarios. Although sensible for the purpose of this comparison, the choice of values does raise some questions for future thought. A superior performance across the calibration scenarios can give a somewhat informative prior, which can adversely affect performance in some scenarios and positively in others.

Both model-based and model-assisted approaches were explored in the study, with differences in their results reflecting the difference in the methods. The model-assisted approaches offer the advantages of fewer assumptions on the dose-response relationship, which is clear to see in very strong performance in scenario D of the simulation study, where the pattern of toxicity risk does not adhere to any standard dose-response model. However, without the assistance of a model to guide the escalation, larger number of patients are on average treated on unsafe doses. This pattern is of course not unique to late onset toxicities, but is potentially accentuated by this.

It is worth noting that in order to compare the designs in a fair manner, we have had to simplify some of their features. For example, the TITE-CRM is capable of including a continuous time-to-event as opposed to the cycle of the event. If the timing of the event is critical for the particular drug in the trial then the impact of this should be given more consideration.

Here we assumed a target toxicity of 0.391 over the three cycles, equivalent to a target of 0.3 in cycle 1 in our framework. This is to remain in line with the standard dose-finding trials that only consider the first cycle, with a target of 0.3, whilst at the same time controlling the rate of DLT across the three cycles. In many of the original proposals, the rate of 0.3 is used as the target for all cycles. It is not expected that such a change would have a substantial effect on the comparison, and such a choice must be made by the clinicians for any individual trial. Although one would have to reconsider the stopping rules that include estimates from cycle 1 too. 

In these implementations we have assumed in our data generation that the risk of DLT decreases linearly across cycles. It may be of interest to additionally perform a sensitivity analysis on this relationship. However, careful consideration must be given to the data generation process and the definition of the target in order to correctly identify the true MTD in each scenario.

The scenarios we investigated were purposefully difficult in order to test the methods. For example, in scenario B, there is only a 0.05 difference in toxicity risk between the 3.5MBq and 4.5MBq dose. This is illustrated with the empirical benchmark only achieving 50\% PCS. Results as such must be considered in the context of the scenario. In a real dose-finding trial, it is likely that the scenario may be `easier' and hence we should not be concerned by such low PCS in this case. 

However, in scenario A, the low PCS in setting 2 across methods is of some concern. Setting 2 is more reflective of the implementation of a dose-finding trial, with stopping rules in place to ensure the safety of patients. In this case these stopping rules are then overly implemented. It is perhaps worth reconsidering that in trials with late onset toxicities, the traditional safety stopping rules may be too strict.

This highlights the importance of considering the impact of the stopping rules on the different statistical methods, as they have a large effect on the true performance of any method used in practice. Whilst it is both interesting and informative to explore the behaviour of the approaches without the implementation of stopping rules, there is a limit to the usefulness of such simulations in isolation.

As well as safety stopping rules, the impact of the precision stopping rule must be considered. This is not applied to the model-assisted methods as the precision of the MTD cannot be estimated without a model. In the case of the nTTP, the precision rule is very often implemented too soon and hence the MTD is underestimated. It is therefore important to consider when using alternative measures of toxicity, whether the traditional approach to stopping rules are actually applicable.
\section{Data Availability Statement}
All data is simulated according to the specifications described.

\section{Acknowledgements}
This report is independent research supported by the National Institute for Health Research (Prof Jaki’s Senior Research Fellowship, NIHR-SRF-2015-08-001) and by the NIHR Cambridge Biomedical Research Centre (BRC-1215-20014). The views expressed in this publication are those of the authors and not necessarily those of the NHS, the National Institute for Health Research or the Department of Health and Social Care (DHSC). T Jaki and H Barnett received funding from UK Medical Research Council (MC\_UU\_00002/14).

\bibliography{dose_finding_cycles}

\begin{table}[ht]
\centering
\begin{tabular}{rrrrrrr}
\multicolumn{1}{c}{}  &\multicolumn{6}{c}{\textit{$p_1$} }  \\
  \hline
 & 1.5MBq & 2.5MBq & 3.5MBq & 4.5MBq & 6.0MBq & 7.0MBq \\ 
  \hline
\textbf{ P.S.1} & \textbf{0.300} & 0.400 & 0.450 & 0.500 & 0.550 & 0.600 \\ 

\textbf{ P.S.2} & 0.050 & 0.070 & 0.100 & 0.150 & 0.200 & \textbf{0.300} \\ 

\textbf{ P.S.3} & 0.100 & 0.200 & \textbf{0.300} & 0.400 & 0.500 & 0.600 \\ 

\textbf{ P.S.4} & 0.150 & 0.200 & 0.250 & \textbf{0.300} & 0.350 & 0.400 \\ 

\textbf{ P.S.5} & 0.400 & 0.450 & 0.500 & 0.550 & 0.600 & 0.650 \\ 

\textbf{ P.S.6} & 0.070 & 0.090 & 0.110 & 0.130 & 0.150 & 0.170 \\ 
   \hline
\end{tabular}
\caption{The P(DLT) in cycle 1 for the six doses in the six scenarios used in the prior callibration procedure, with the MTD highlighted in \textbf{boldface}.\label{tab:prior_sc}}
\end{table}

\begin{table}[ht!]
\centering
\begin{tabular}{@{}lll@{}}
\toprule
                    & \textbf{Setting 1}                              & \textbf{Setting 2}                              \\ \midrule
\textbf{TITE-CRM}    & $\sigma^2=1$                                    & $\sigma^2=1$                                    \\
                    & $d=(0.05, 0.10, 0.15, 0.20, 0.25, 0.30)$        & $d=(0.05, 0.10, 0.15, 0.20, 0.25, 0.30)$     \\ \midrule
                    & $ESS \approx 2$ &  $ESS \approx 2$                 \\ \midrule
\textbf{TITE-CRM2}   & $\mu_{a_0}=-1$                                      & $\mu_{a_0}=-1$                                      \\
                    & $\sigma_{a_0}^{-2}=0.3$                             & $\sigma_{a_0}^{-2}=0.3$                             \\
                    & $\mu_{a_1}=log(0.2)$                                & $\mu_{a_1}=log(0.2)$                                \\
                    & $\sigma_{a_1}^{-2}=0.3$                             & $\sigma_{a_1}^{-2}=0.3$                             \\ \midrule
                    & $ESS \approx 1$ &  $ESS \approx 1$               \\ \midrule
  
\textbf{ICSDP}      & $\pi*_1=0.2$                                    & $\pi*_1=0.2$                                    \\
                    & $\pi*_J=0.4$                                    & $\pi*_J=0.3$                                    \\
                    & $n_0=6$                                         & $n_0=4$                                         \\ \midrule
                    & $ESS \approx 2$ &  $ESS \approx 1$               \\ \midrule
\textbf{POMM}       & $p_1*=(0.15, 0.20, 0.25, 0.3, 0.35, 0.40)$      & $p_1*=(0.15, 0.20, 0.25, 0.3, 0.35, 0.40)$      \\
                    & $n_0=2$                                         & $n_0=2$                                         \\
                    & $p^{G2}_1/p_1 = (0.20, 0.30, 0.40, 0.50, 0.60)$ & $p^{G2}_1/p_1 = (0.20, 0.30, 0.40, 0.50, 0.60)$ \\ \midrule
                    & $ESS \approx 2$ &  $ESS \approx 2$               \\ \midrule
\textbf{nTTP}       & $\mu_{\beta_0}=0.1$                             & $\mu_{\beta_0}=0.05$                            \\
                    & $\sigma^2_{\beta_0}=100$                        & $\sigma^2_{\beta_0}=10$                         \\
                    & $\mu_{\beta_1}=0.5$                             & $\mu_{\beta_1}=0.1$                             \\
                    & $\sigma^2_{\beta_1}=100$                        & $\sigma^2_{\beta_1}=10$                         \\
                    & $\mu_{\beta_2}=0$                               & $\mu_{\beta_2}=0$                               \\
                    & $\sigma^2_{\beta_2}=10$                         & $\sigma^2_{\beta_2}=10$                         \\ \midrule
                    & $ESS \approx 1$ &  $ESS \approx 1$               \\ \midrule
\textbf{TITE-BOIN}  & $\tau_1=0.3128$                                (equivalent to $\lambda_e=0.3512$) & $\tau_1=0.3128$                                (equivalent to $\lambda_e=0.3512$) \\
                    & $\tau_2=0.5083$                                 (equivalent to $\lambda_d=0.4492$)                        & $\tau_2=0.5083$         (equivalent to $\lambda_d=0.4492$)                        \\
                    & $\alpha=0.1$                                    & $\alpha=1$                                      \\
                    & $\beta=0.9$                                     & $\beta=1$                                       \\ \midrule
                    & $ESS \approx 1$ &  $ESS \approx 2$               \\ \midrule
\textbf{TITE-mTPI2} & $\tau_1=0.3519$                                 & $\tau_1=0.3519$                                 \\
                    & $\tau_2=0.5474$                                 & $\tau_2=0.5474$                                 \\ \midrule
                    & $ESS \approx 2$ &  $ESS \approx 2$               \\ \midrule
\textbf{R-mTPI2}    & $\tau_1=0.3519$                                 & $\tau_1=0.3519$                                 \\
                    & $\tau_2=0.5474$                                 & $\tau_2=0.5474$                                 \\ \midrule
                    & $ESS \approx 2$ &  $ESS \approx 2$               \\ 
                    \bottomrule
\end{tabular}
\caption{The values of hyper-parameters resulting from the prior callibration procedure. Note that this corresponds to a target toxicity of 0.391 over 3 cycles. $ESS$ is the average effective sample size per dose level. \label{tab:prior_par}}
\end{table}

\begin{table}[ht]
\centering
\small{
\begin{tabular}{ccccccccc}
  \hline
   & \rotatebox{60}{\textbf{TITE-CRM}} & \rotatebox{60}{\textbf{TITE-CRM2}} & \rotatebox{60}{\textbf{ICSDP}} & \rotatebox{60}{\textbf{POMM}} & \rotatebox{60}{\textbf{nTTP}} & \rotatebox{60}{\textbf{TITEBOIN}} & \rotatebox{60}{\textbf{TITE-mTPI2}} & \rotatebox{60}{\textbf{R-mTPI2}}  \\ 
 \hline 
  \textit{Scenario} & \multicolumn{8}{c}{\textit{Mean Duration in Weeks (sd) }}\\
  \hline
\textbf{A} & 39 (11) & 36 (10) & 46 (10) & 52 (9) & 32 (5) & 40 (14) & 46 (23) & 50 (26) \\ 
  \textbf{B} & 55 (14) & 53 (15) & 55 (10) & 55 (8) & 44 (14) & 57 (13) & 81 (31) & 95 (34) \\ 
  \textbf{C} & 34 (9) & 32 (7) & 44 (10) & 52 (9) & 31 (4) & 35 (16) & 37 (22) & 37 (23) \\ 
  \textbf{D} & 63 (7) & 57 (8) & 57 (5) & 55 (6) & 56 (10) & 49 (3) & 85 (7) & 100 (9) \\ 
  \textbf{Mean} & 48 & 45 & 51 & 53 & 41 & 45 & 62 & 70 \\ 
   \hline
  \textit{Scenario} & \multicolumn{8}{c}{ \textit{ Mean Number of Patients (sd)} }\\
  \hline
    \textbf{A} & 14 (6) & 12 (5) & 17 (5) & 20 (4) & 10 (2) & 19 (6) & 15 (5) & 15 (5) \\ 
  \textbf{B} & 22 (7) & 21 (8) & 22 (5) & 22 (4) & 16 (7) & 26 (4) & 21 (6) & 23 (7) \\ 
  \textbf{C} & 11 (4) & 10 (3) & 16 (5) & 20 (4) & 10 (2) & 17 (7) & 13 (5) & 12 (5) \\ 
  \textbf{D} & 26 (4) & 23 (4) & 23 (3) & 22 (2) & 23 (5) & 24 (1) & 24 (2) & 25 (2) \\ 
  \textbf{Mean} & 18 & 17 & 20 & 21 & 15 & 22 & 18 & 19 \\ 
   \hline

\end{tabular}}
\caption{Setting 1: Measures of size of the trial across scenarios, total duration in weeks and total number of patients. \label{tab:size_stage1}}
\end{table}

\begin{table}[ht]
\centering
\small{
\begin{tabular}{ccccccccc}
  \hline
   & \rotatebox{60}{\textbf{TITE-CRM}} & \rotatebox{60}{\textbf{TITE-CRM2}} & \rotatebox{60}{\textbf{ICSDP}} & \rotatebox{60}{\textbf{POMM}} & \rotatebox{60}{\textbf{nTTP}} & \rotatebox{60}{\textbf{TITEBOIN}} & \rotatebox{60}{\textbf{TITE-mTPI2}} & \rotatebox{60}{\textbf{R-mTPI2}} \\ 
  \hline

  \textit{Stopping Reason} & \multicolumn{8}{c}{ \textit{ Scenario A} }\\
  \hline
\textbf{Sufficient Information} & 99 & 100 & 100 & 97 & 100 & 95 & 100 & 99 \\ 
  \textbf{Maximum Patients} & 2 & 1 & 2 & 3 & 0 & 10 & 1 & 1 \\ 
    \hline
  \textit{Stopping Reason} & \multicolumn{8}{c}{ \textit{ Scenario B} }\\
  \hline
  
  \textbf{Sufficient Information} & 90 & 88 & 95 & 95 & 96 & 80 & 95 & 75 \\ 
  \textbf{Maximum Patients} & 21 & 22 & 11 & 5 & 4 & 41 & 11 & 32 \\ 
      \hline
  \textit{Stopping Reason} & \multicolumn{8}{c}{ \textit{ Scenario C }}\\
  \hline
  \textbf{Sufficient Information} & 100 & 100 & 100 & 96 & 100 & 95 & 100 & 100 \\ 
  \textbf{Maximum Patients} & 1 & 0 & 1 & 4 & 0 & 8 & 1 & 1 \\ 
      \hline
  \textit{Stopping Reason} & \multicolumn{8}{c}{ \textit{ Scenario D} }\\
  \hline
  \textbf{Sufficient Information} & 89 & 94 & 100 & 99 & 96 & 99 & 99 & 99 \\ 
  \textbf{Maximum Patients} & 32 & 16 & 2 & 1 & 4 & 2 & 2 & 5 \\ 
   \hline
\end{tabular}}
\caption{Setting 1: Summary of stopping reasons for setting 1, expressed as a percentage of simulations where the given stopping rule was triggered. The sum of these may be greater than 100, since it is possible for more than one rule to be triggered in a single trial. \label{tab:stopping_stage1}}
\end{table}

\begin{table}[ht]
\centering
\small{
\begin{tabular}{ccccccccc}
  \hline
   & \rotatebox{60}{\textbf{TITE-CRM}} & \rotatebox{60}{\textbf{TITE-CRM2}} & \rotatebox{60}{\textbf{ICSDP}} & \rotatebox{60}{\textbf{POMM}} & \rotatebox{60}{\textbf{nTTP}} & \rotatebox{60}{\textbf{TITEBOIN}} & \rotatebox{60}{\textbf{TITE-mTPI2}} & \rotatebox{60}{\textbf{R-mTPI2}}  \\ 
 \hline 
  \textit{Scenario} & \multicolumn{8}{c}{\textit{Mean Duration in Weeks (sd) }}\\
  \hline
\textbf{A} & 39 (13) & 31 (13) & 44 (12) & 44 (13) & 29 (4) & 33 (18) & 41 (26) & 43 (30) \\ 
  \textbf{B} & 52 (12) & 50 (16) & 53 (10) & 45 (9) & 32 (6) & 51 (17) & 78 (32) & 90 (38) \\ 
  \textbf{C} & 32 (13) & 25 (11) & 38 (14) & 39 (15) & 27 (5) & 27 (19) & 32 (25) & 29 (27) \\ 
  \textbf{D} & 57 (9) & 50 (10) & 49 (5) & 36 (7) & 29 (5) & 40 (7) & 76 (10) & 92 (12) \\ 
  \textbf{Mean} & 45 & 39 & 46 & 41 & 29 & 38 & 57 & 63 \\ 
      \hline
  \textit{Scenario} & \multicolumn{8}{c}{ \textit{ Mean Number of Patients (sd)} }\\
  \hline
  \textbf{A} & 14 (6) & 10 (6) & 16 (6) & 17 (6) & 9 (2) & 16 (8) & 13 (7) & 12 (7) \\ 
  \textbf{B} & 20 (6) & 19 (8) & 21 (5) & 17 (4) & 10 (3) & 24 (7) & 20 (7) & 22 (8) \\ 
  \textbf{C} & 11 (6) & 8 (5) & 14 (6) & 14 (7) & 8 (2) & 13 (9) & 10 (7) & 9 (6) \\ 
  \textbf{D} & 23 (5) & 20 (4) & 19 (3) & 13 (3) & 10 (2) & 20 (3) & 20 (3) & 21 (3) \\ 
  \textbf{Mean} & 17 & 14 & 17 & 15 & 9 & 18 & 16 & 16 \\ 
  \hline
\end{tabular}}
\caption{Setting 2: Measures of size of the trial across scenarios, total duration in weeks and total number of patients. \label{tab:size_stage2}}
\end{table}

\begin{table}[ht]
\centering
\small{
\begin{tabular}{ccccccccc}
  \hline
 & \rotatebox{60}{\textbf{TITE-CRM}} & \rotatebox{60}{\textbf{TITE-CRM2}} & \rotatebox{60}{\textbf{ICSDP}} & \rotatebox{60}{\textbf{POMM}} & \rotatebox{60}{\textbf{nTTP}} & \rotatebox{60}{\textbf{TITEBOIN}} & \rotatebox{60}{\textbf{TITE-mTPI2}} & \rotatebox{60}{\textbf{R-mTPI2}} \\ 
  \hline

  \textit{Stopping Reason} & \multicolumn{8}{c}{ \textit{ Scenario A} }\\
  \hline

Sufficient Information & 57 & 46 & 88 & 60 & 46 & 68 & 70 & 68 \\ 
  Lowest Dose Deemed Unsafe & 42 & 50 & 11 & 20 & 22 & 28 & 30 & 32 \\ 
  Highest Dose Deemed too Safe & 0 & 0 & 0 & 0 & 0 & 0 & 0 & 0 \\ 
  Precision & 0 & 17 & 2 & 14 & 73 & 0 & 0 & 0 \\ 
  Hard Safety & 10 & 4 & 11 & 14 & 10 & 3 & 4 & 4 \\ 
  Maximum Patients & 2 & 1 & 1 & 2 & 0 & 7 & 1 & 1 \\ 
  Unsafe (Total) & 42 & 50 & 11 & 26 & 25 & 28 & 30 & 32 \\ 
  Sufficient/Precision (Total) & 58 & 50 & 89 & 71 & 75 & 68 & 70 & 68 \\ 
      \hline
  \textit{Stopping Reason} & \multicolumn{8}{c}{ \textit{ Scenario B} }\\
  \hline
  Sufficient Information & 65 & 65 & 83 & 35 & 32 & 77 & 89 & 70 \\ 
  Lowest Dose Deemed Unsafe & 4 & 11 & 1 & 1 & 1 & 6 & 6 & 8 \\ 
  Highest Dose Deemed too Safe & 0 & 1 & 0 & 0 & 0 & 2 & 1 & 1 \\ 
  Precision & 34 & 21 & 21 & 72 & 96 & 0 & 0 & 0 \\ 
  Hard Safety & 1 & 0 & 1 & 1 & 1 & 0 & 0 & 0 \\ 
  Maximum Patients & 7 & 17 & 9 & 2 & 0 & 29 & 8 & 27 \\ 
  Unsafe (Total) & 4 & 11 & 1 & 2 & 1 & 6 & 6 & 8 \\ 
  Sufficient/Precision (Total) & 94 & 79 & 95 & 97 & 98 & 77 & 89 & 70 \\ 
      \hline
  \textit{Stopping Reason} & \multicolumn{8}{c}{ \textit{ Scenario C} }\\
  \hline
  Sufficient Information & 37 & 24 & 70 & 42 & 20 & 49 & 49 & 47 \\ 
  Lowest Dose Deemed Unsafe & 63 & 73 & 29 & 38 & 71 & 48 & 51 & 53 \\ 
  Highest Dose Deemed too Safe & 0 & 0 & 0 & 0 & 0 & 0 & 0 & 0 \\ 
  Precision & 0 & 19 & 1 & 6 & 32 & 0 & 0 & 0 \\ 
  Hard Safety & 24 & 10 & 29 & 30 & 21 & 9 & 10 & 10 \\ 
  Maximum Patients & 1 & 0 & 1 & 2 & 0 & 6 & 1 & 0 \\ 
  Unsafe (Total) & 63 & 73 & 29 & 50 & 72 & 48 & 51 & 53 \\ 
  Sufficient/Precision (Total) & 37 & 27 & 71 & 48 & 28 & 49 & 49 & 47 \\ 
      \hline
  \textit{Stopping Reason} & \multicolumn{8}{c}{ \textit{ Scenario D} }\\
  \hline
  Sufficient Information & 96 & 64 & 63 & 5 & 8 & 99 & 99 & 99 \\ 
  Lowest Dose Deemed Unsafe & 0 & 1 & 0 & 0 & 0 & 1 & 1 & 1 \\ 
  Highest Dose Deemed too Safe & 0 & 0 & 0 & 0 & 0 & 0 & 0 & 0 \\ 
  Precision & 0 & 44 & 60 & 99 & 100 & 0 & 0 & 0 \\ 
  Hard Safety & 0 & 0 & 0 & 0 & 0 & 0 & 0 & 0 \\ 
  Maximum Patients & 13 & 4 & 0 & 0 & 0 & 1 & 1 & 1 \\ 
  Unsafe (Total) & 0 & 1 & 0 & 0 & 0 & 1 & 1 & 1 \\ 
  Sufficient/Precision (Total) & 96 & 98 & 100 & 100 & 100 & 99 & 99 & 99 \\ 
   \hline
\end{tabular}}
\caption{Setting 2: Summary of stopping reasons for setting 2, expressed as a percentage of simulations where the given stopping rule was triggered. The sum of these may be greater than 100, since it is possible for more than one rule to be triggered in a single trial. \label{tab:stopping_stage2}}
\end{table}

\end{document}